%% file: LifeACRI.tex
\documentclass{llncs}

\usepackage[latin1]{inputenc}
\usepackage{epsfig}
\usepackage{tabularx}
\include{commands}

\newcommand{\SzA}{5.5}
\newcommand{\szEndFig}{5.6}
\newcommand{\Life}{{\it Life}}

\begin{document}

\title{ Perturbing the Topology of the Game of Life Increases its Robustness to	Asynchrony }
\titlerunning{Perturbing the Game of Life Increases its Robustness to	Asynchrony}
\author{ Nazim Fatès \inst{1} \and Michel Morvan  \inst{1,2} }
\institute{	 ENS Lyon - LIP \\ UMR CNRS - ENS Lyon - UCB Lyon - INRIA 5668 \\ 
	 46, allée d'Italie - 69 364 Lyon Cedex 07 - France.\\
\and
	Institut Universitaire de France\\
	\email{ \{Nazim.Fates\},\{Michel.Morvan\}@ens-lyon.fr} 
}
	
\maketitle
\begin{center}
	
\end{center}

\begin{abstract}
An experimental analysis of the asynchronous version of the ``Game of Life'' 
is performed to estimate how topology perturbations modify its evolution.
We focus on the study of a phase transition from an ``inactive-sparse phase'' to  a ``labyrinth phase''  and 
produce experimental data to quantify these changes 
as a function of the density of the initial configuration, the value of the synchrony rate, and the topology missing-link rate. 
An interpretation of the experimental results is given using the hypothesis that initial ``germs'' colonize the whole lattice 
and the validity of this hypothesis is tested.
\end{abstract}

\section{Introduction}

Cellular automata were originally introduced by von Neumann in order to study the logical properties of self-reproducing machines.
Following Ulam's suggestions, the requirements he made for constructing such a machine was 
the discreteness of space using cells, 
discreteness of time using an external clock, 
the symmetry of the rules governing cells interaction, 
and the locality of these interactions; it resulted in the birth of the cellular automaton (CA) model.
In order to make the self-reproduction not trivial he also required that the self-reproducing machine should be 
computation-universal (e.g., \cite{Pou85}).
The resulting CA used 29 elementary states for each cell and updates used 5 neighbors.
Later on, Conway introduced a CA called ``Game of Life'' or simply \Life\ 
which was also proved to be computation universal \cite{Ber82}. 
This CA is simpler than von Neumann's in at least two ways: 
the local rule uses only two states and it can be summarized with by sub-rules (birth and death rules).

However, the question remained open to know what is the importance of perfect synchrony on a CA behavior.
Indeed, since the first study on the effects of asynchronous update carried out by Ingerson and Buvel \cite{Ingerson84}, 
many criticisms have been addressed to the use of the CA as models of natural phenomena.
Some authors investigated, using various techniques, how synchrony variations changed CA qualitative behavior \cite{Ingerson84}, 
\cite{Huberman93}, \cite{Ber94}, \cite{Blo99}, \cite{Sch99}, \cite{Fat04}. 
All studies agree on the fact that for some CA, there are situations in which small changes in the update method lead to 
qualitative changes of the evolution of the CA, thus showing the need for further studies of robustness to asynchronism.
Similarly, some authors investigated the effect of perturbing the topology (i.e., the links between cells) 
in one dimension \cite{Ila01} by adding links, 
or in two dimensions with small-world construction algorithms \cite{Ser02}, \cite{Hua03}. 
Here too, the studies showed that robustness to topology changes was a key factor in the CA theory and 
that some CA showed ``phase transitions'' when varying the intensity of the topology perturbation.

The aim of this work is to question, in the case of \Life, the importance of the two hypotheses used in the classical CA paradigm: 
what happens when the CA is no longer perfectly synchronous and when the topology is perturbed?
In Section \ref{Sec:Model}, we present the model and describe the qualitative behavior 
induced by the introduction of asynchronism and/or topology perturbations. 
In Section \ref{Sec:Qualitative}, we observe that (i) \Life\ is sensitive to asynchronism; (ii) robust to topology perturbations 
and (iii) that the robustness to asynchronism is increased when the topology characteristics become irregular. 
Section \ref{Sec:ExperimentalData} is devoted to presenting a rigorous experimental validation 
and exploration of these phenomena 
for which a potential explanation 
based on the notion of ``germ'' development is discussed and studied in Section \ref{Sec:Analysis}.

\section{The Model} \label{Sec:Model}

Classically, \Life\  is run on a regular subset of $ \Z^2 $. 
For simulation purposes, 
the configurations are finite squares with $ N \times N $ cells and the neighborhood of each cell 
is constituted of the cell itself and the 8 nearest neighbors (Moore neighborhood).  
We use periodic boundary conditions meaning that all cell position indices are taken in $ {\Z}/{N \Z} $. 
The type of boundary conditions does play an important role at least for small configurations as shown in \cite{Blo97}.

\Life\  belongs to the outer-totalistic (e.g., \cite{pac85}, \cite{Ila01}) class of CA: 
the local transition rule $ f $ is specified 
as a function of the present state $ q(c) $ and 
of the number $ S_1(c) $ of cells with states $ \stone$ in the neighborhood.
The \Life\  transition function $ f ( q , S_1 ) $ can be written:
\begin{equation}
	f ( \stzero, S_1 ) = 	1  \text{ if } S_1 = 3 \\  \text{ ; }	f ( \stzero, S_1 ) = 0 \text{ otherwise,}
\tag{birth rule}
\end{equation}
\begin{equation}
	f ( \stone, S_1 ) = 	1  \text{ if } S_1 = 2 \text{ or } S_1 = 3 {;}	 	f ( \stone, S_1 ) = 0 \text{ otherwise.}
\tag{death rule}
\end{equation}

In the sequel, we consider \Life\ as an asynchronous cellular automaton (ACA) acting on a possibly perturbed topology.

There are several asynchronous dynamics: 
one may, for example update cells one by one in a fixed order from left to right and from bottom to top. 
This update method is called ``line-by-line sweep'' \cite{Sch99} and it has been shown that this type of dynamics 
introduce spurious behaviors due to the correlation 
between the spatial arrangement of the cells and the spatial ordering of the updates.
These correlations can only be suppressed with a random updating of the cells.
In this work, we choose to examine only one type of asynchronism which consists in 
applying the local rule, for each cell independently, with probability $ \sr $.
The parameter $ \sr $ is called the ``synchronicity'' \cite{Blo99} or the {\it synchrony rate}; 
one can also view it as a parameter that would control the evolution of a probabilistic cellular automata (PCA) 
where the transition function results in applying \Life\ rule with a probability $ \sr $ and the identity rule with probability $ 1 - \sr $.

We choose to perturb topology by definitely removing links between cells.
Let $ G_0= (\U, E_0) $ be the oriented graph that represents cells interactions: 
$ (c,c') \in E $ if and only if $ c' $ belongs to neighborhood of $ c$.
The graph with perturbed topology $ G=  (\U, E)$ is obtained by examining each cell $ c \in \U $ and, for each cell in the neighborhood of $ c $
and removing the link $ (c, c') $ with a probability $ \er $;
the parameter $ \er $ is called {\em missing-link rate}.
Note that, as the local function is expressed in an outer-totalistic mode, 
we can still apply it on neighborhoods of various sizes. 
The definition we use induces an implicit choice of behavior in the case where a link is missing :
the use of $ S_1 $ in the local rule definition implies that the cell will consider missing cells of the neighborhood as being in state $ \stzero $.
Other choices would have been possible; for example assuming this state to be $ \stone $ or the current value of the cell itself.

\section{Observations and Measures} \label{Sec:Experiments}

\begin{figure}[htbp]
\begin{center}
\begin{tabular}{c c c }
\Orbit{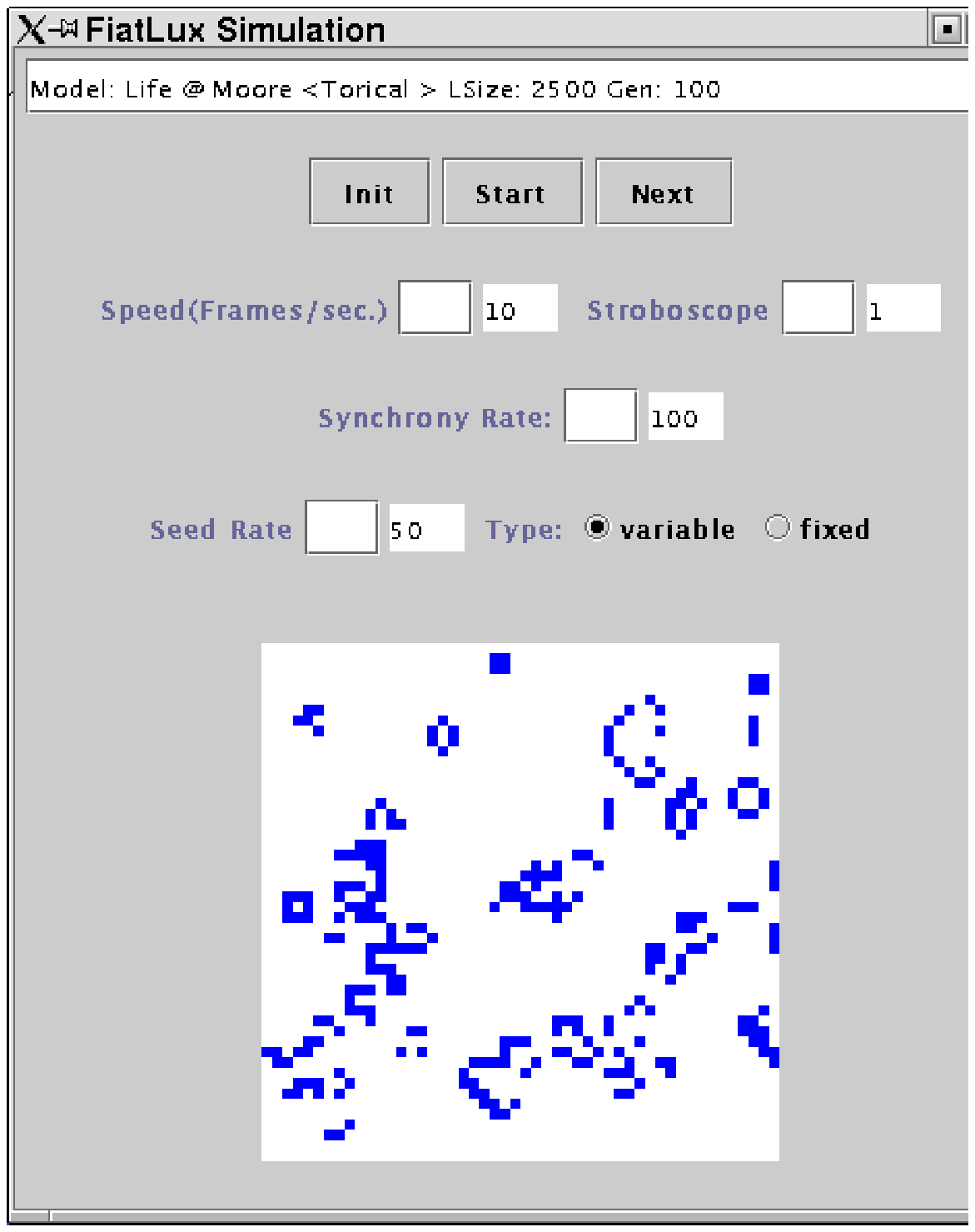} & \Orbit{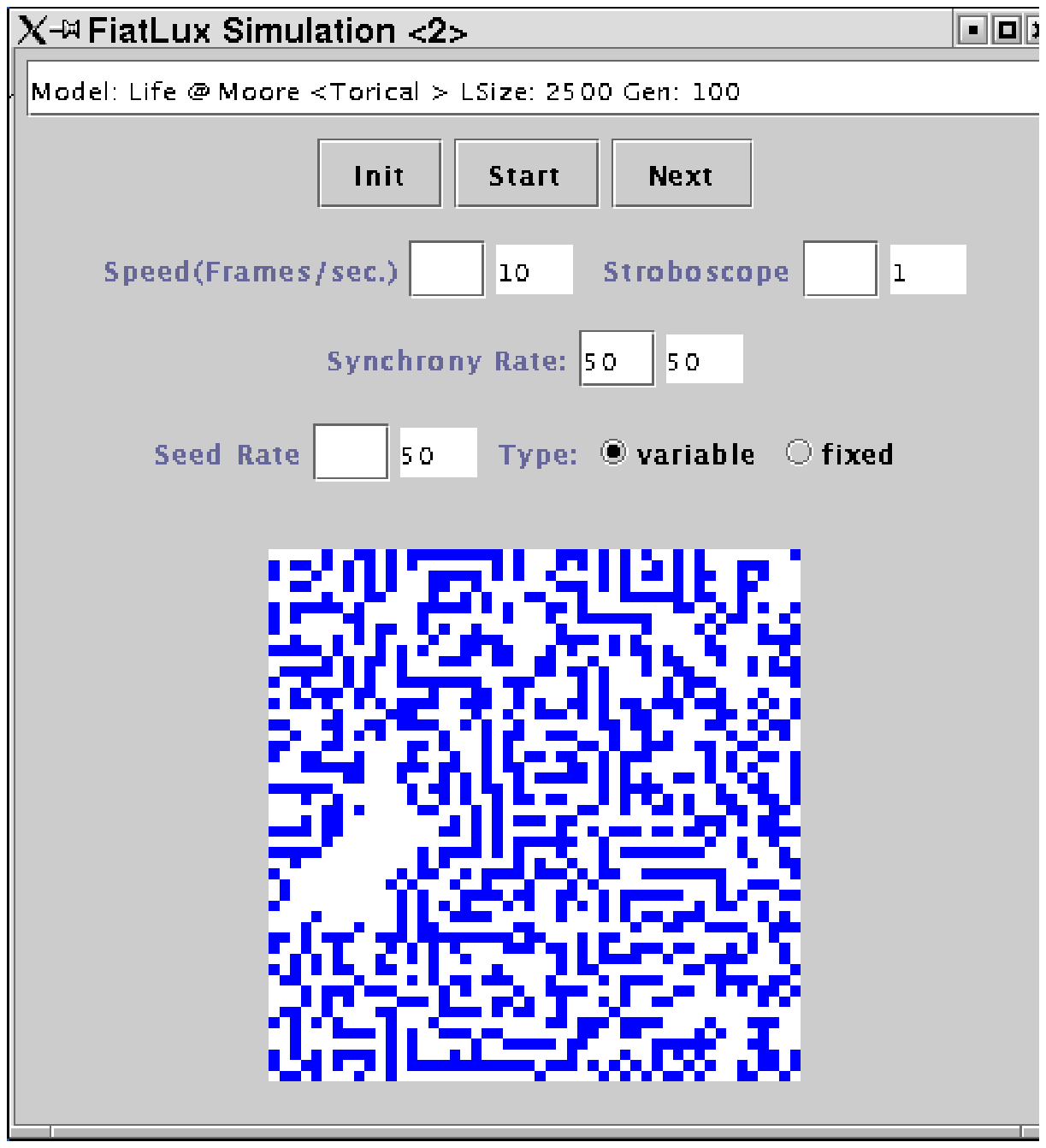} &\Orbit{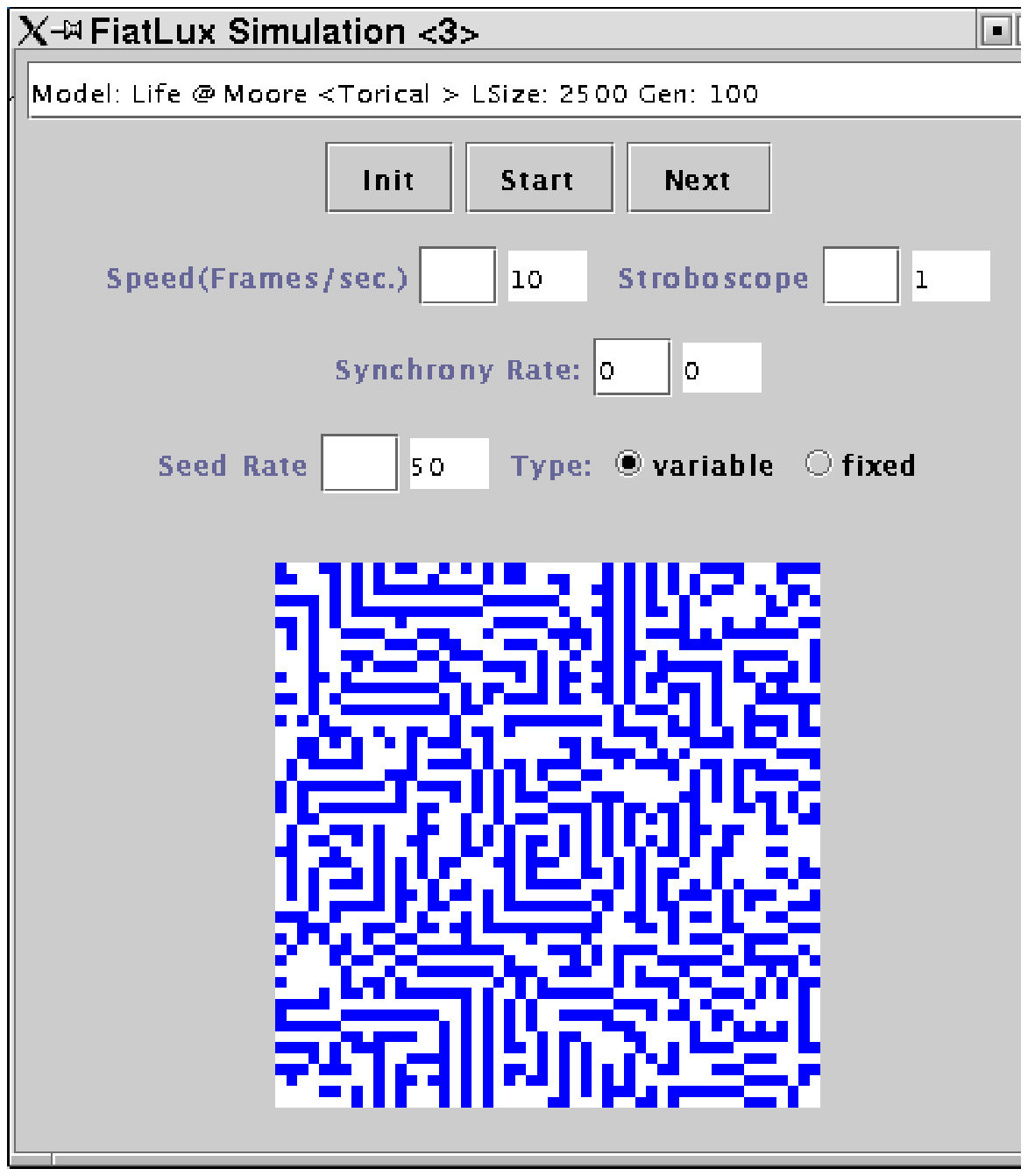} \\
 $ \sr = 1.0 $ &  $ \sr =0.5 $ &  sequential updating \\
\end{tabular}
\caption{\Life\ configurations for $ N=50 \times 50 $, after $ T=100 $ time steps, starting from a random configuration of density $ \dini=0.5 $.
In the sequential updating, cells are randomly updated one after another.
}
\label{Fi:Life1}
\vspace{0.6cm}
\begin{tabular}{ c c c }
\Orbit{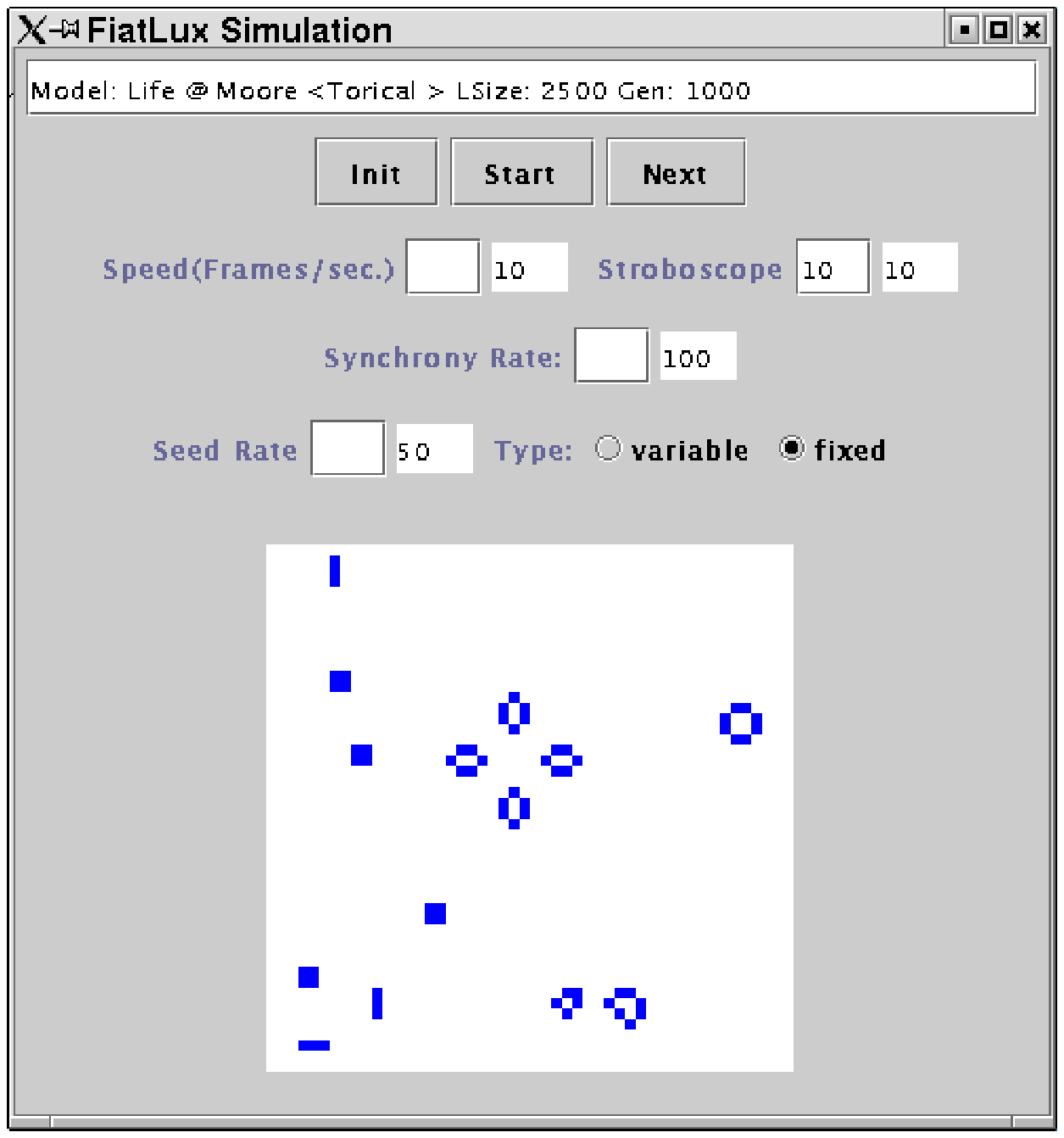} & \Orbit{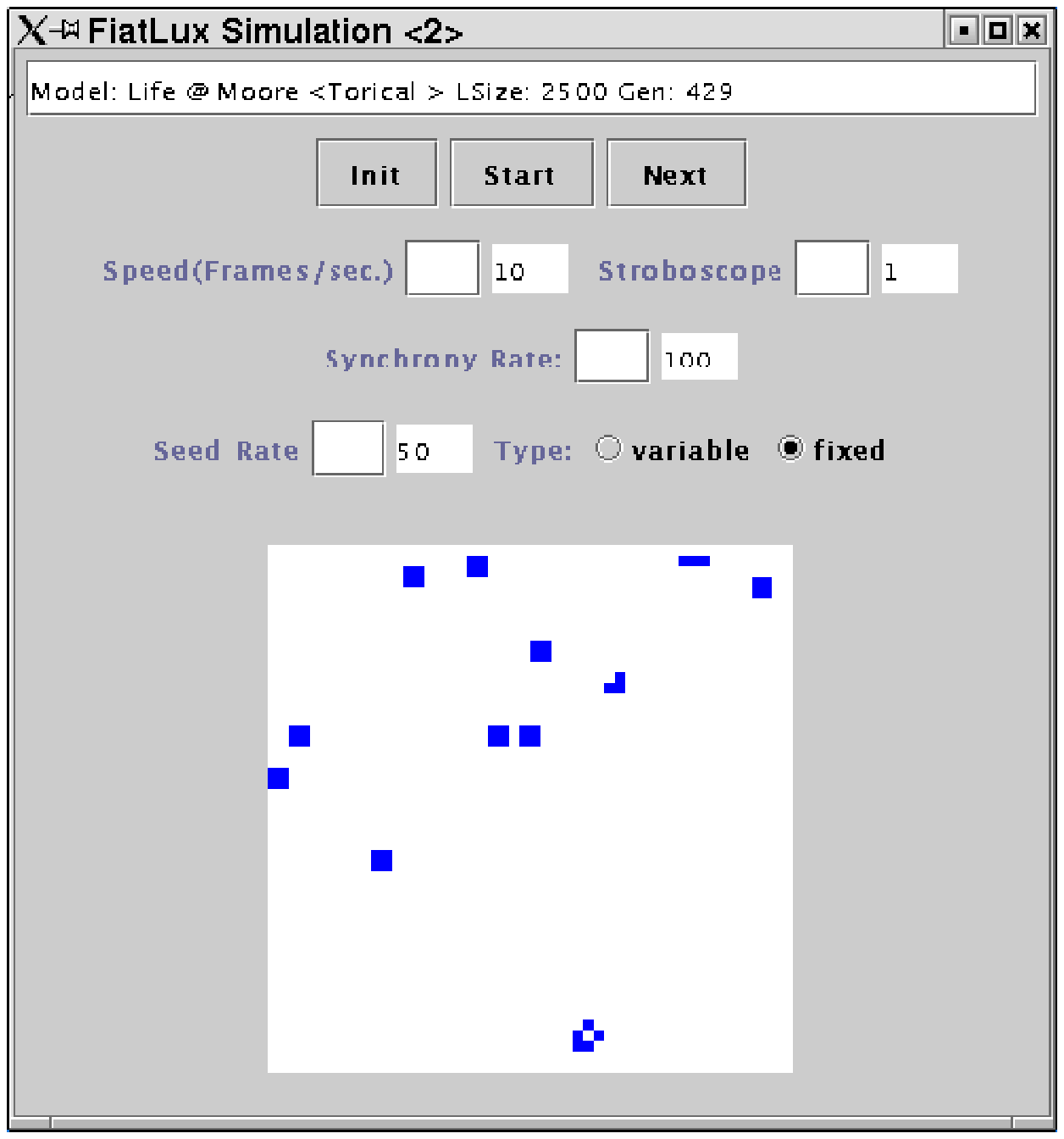} & \Orbit{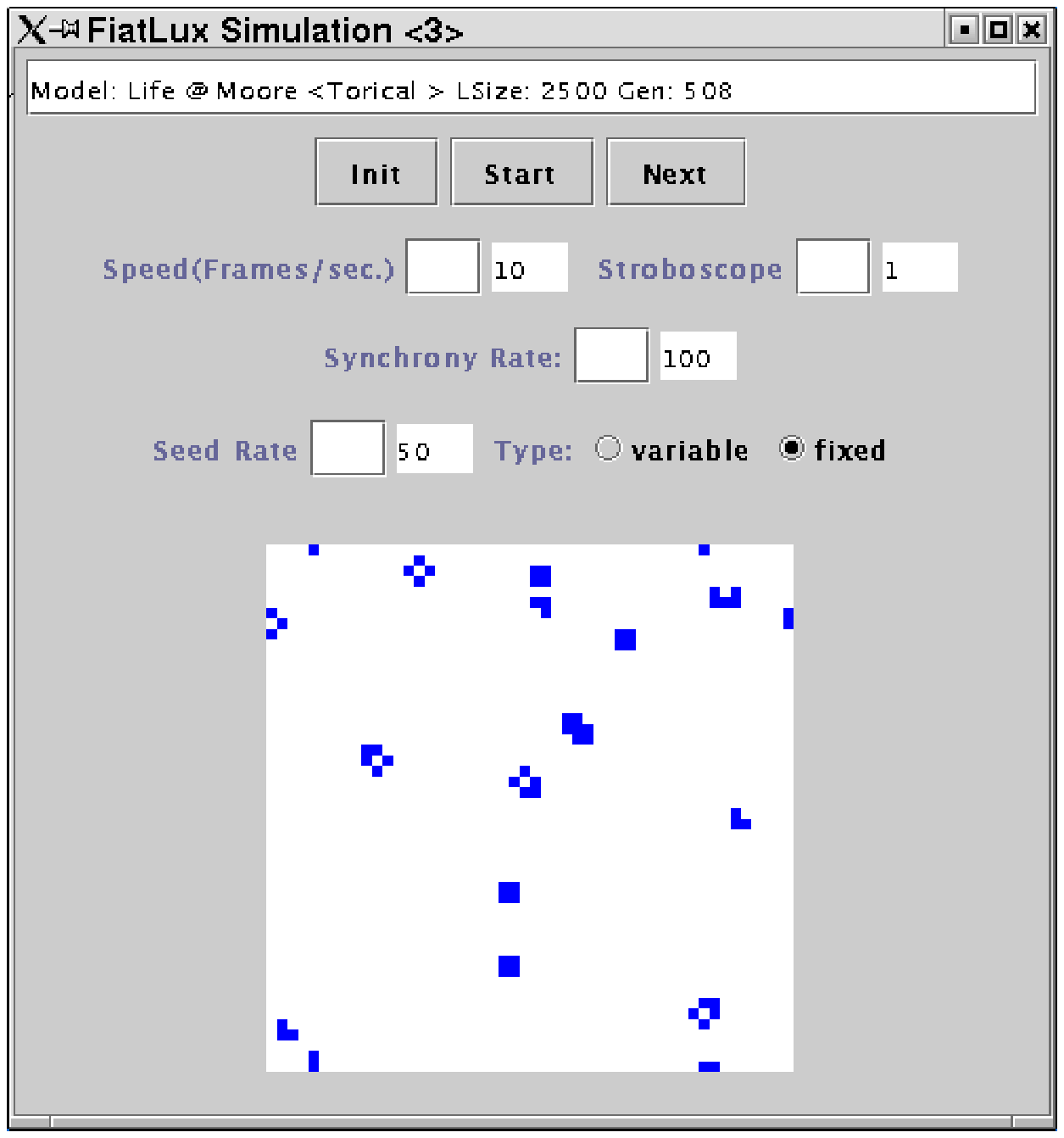} \\
 $ \er = 0 $ & $ \er = 0.05 $  & $ \er = 0.10 $  \\
\end{tabular}
\caption{\Life\ configurations for synchronous evolution ($ \sr= 1.00 $) with $ N=50 \times 50 $, after $ T=1000 $ time steps, starting from a random configuration of density $ \dini=0.5 $.
}
\label{Fi:ALife2}
\vspace{0.6cm}
\begin{tabular}{ c c c}
 \Orbit{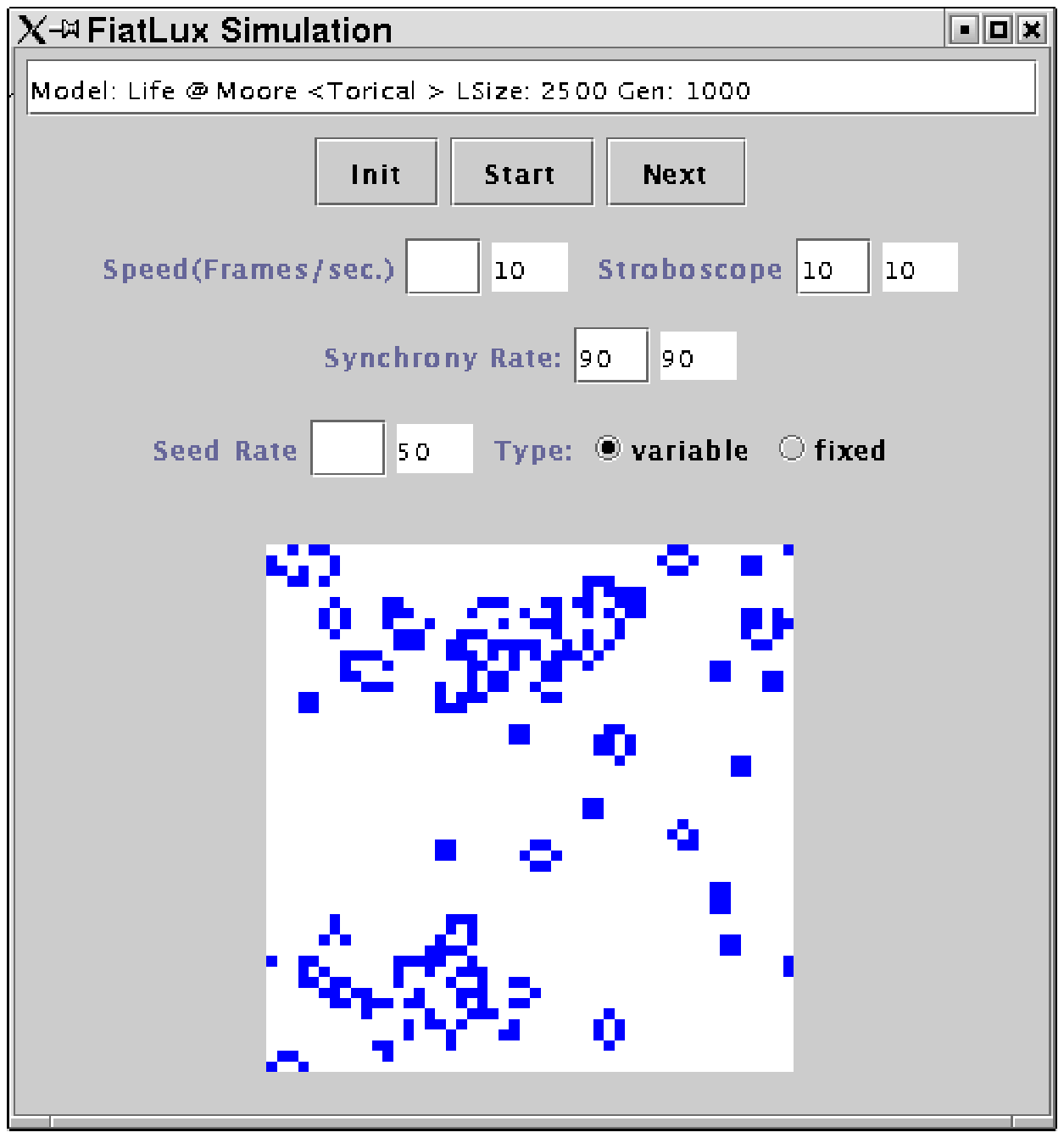} & \Orbit{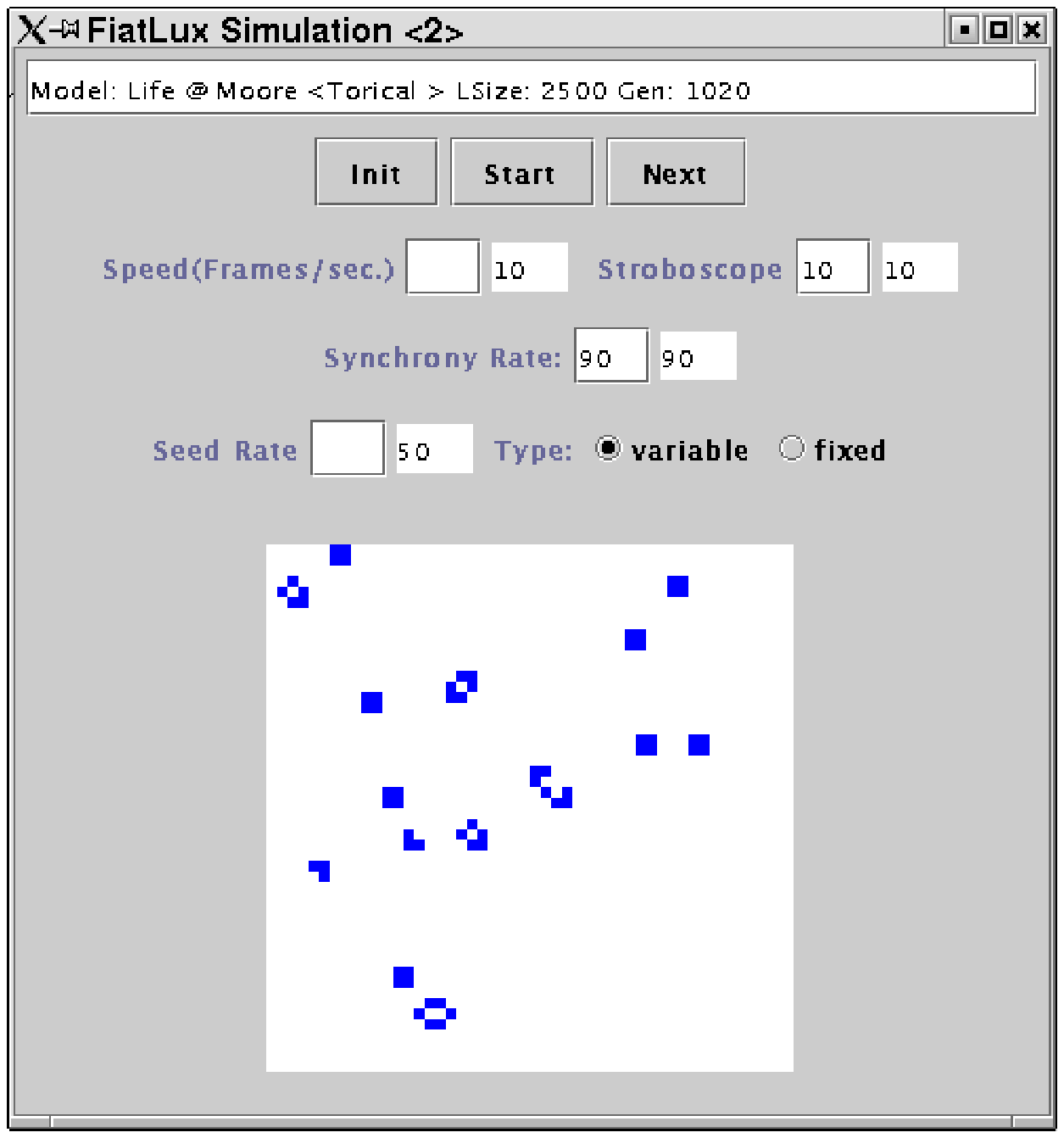} & \Orbit{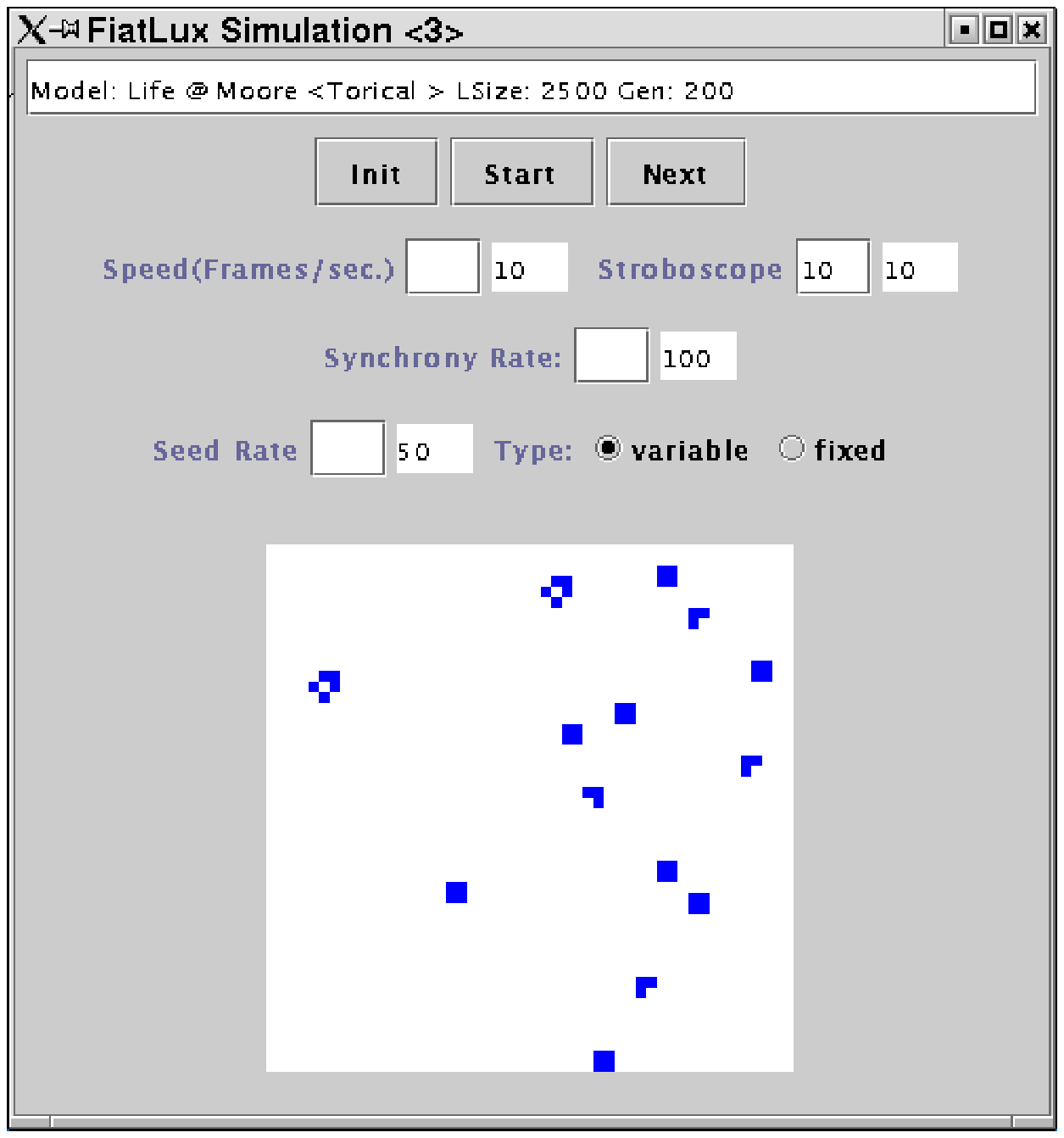}\\
 \Orbit{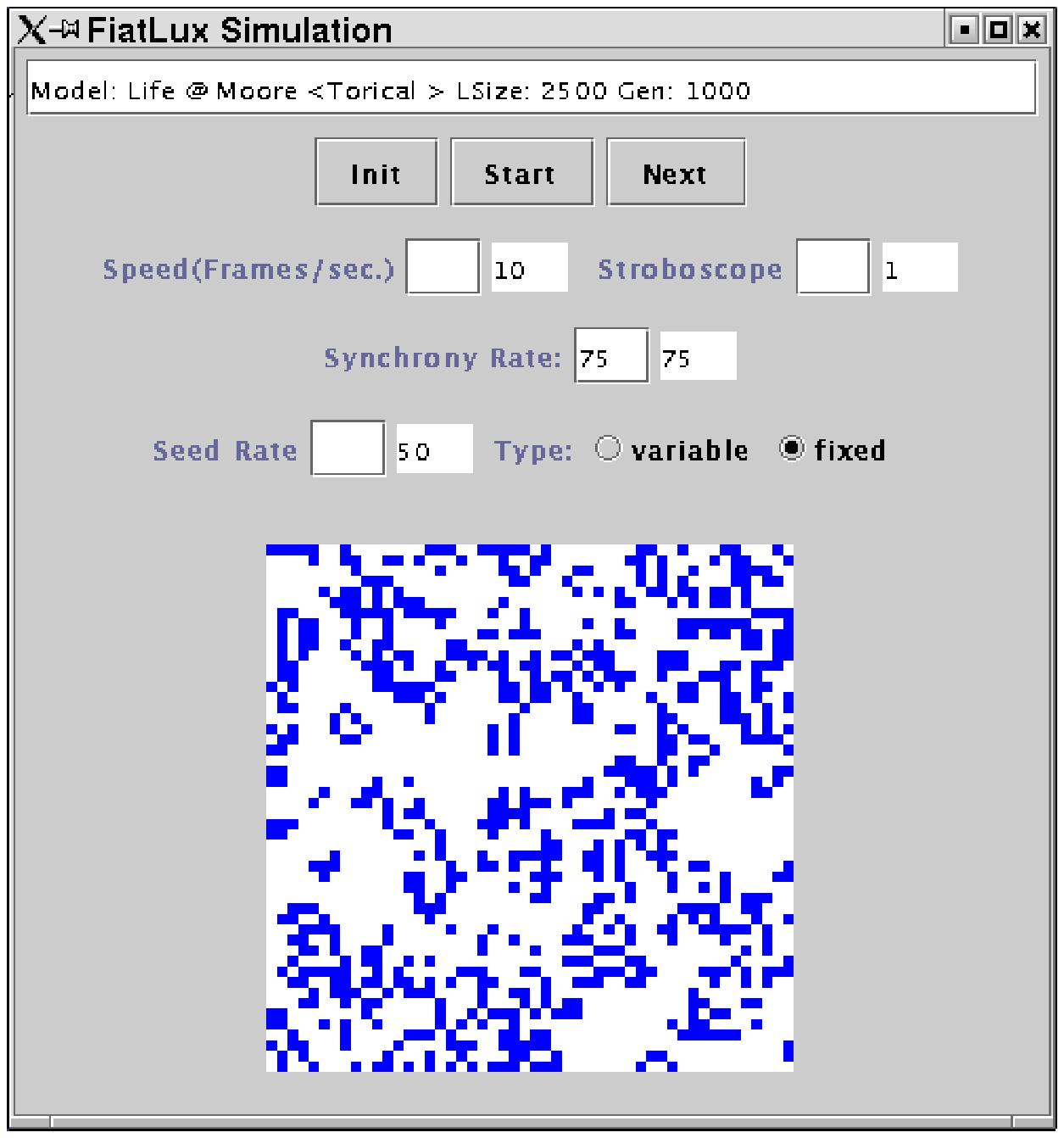} & \Orbit{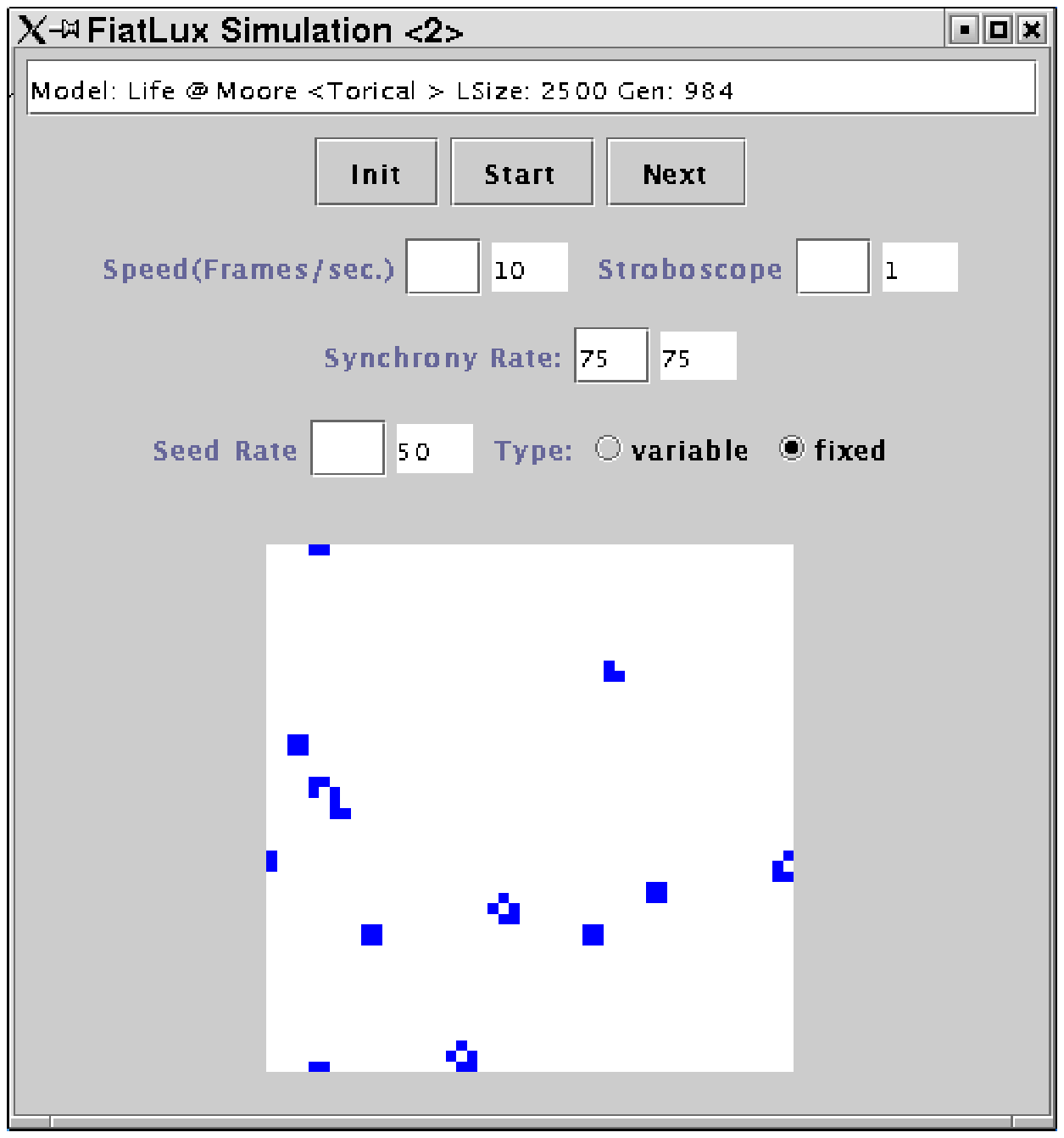} & \Orbit{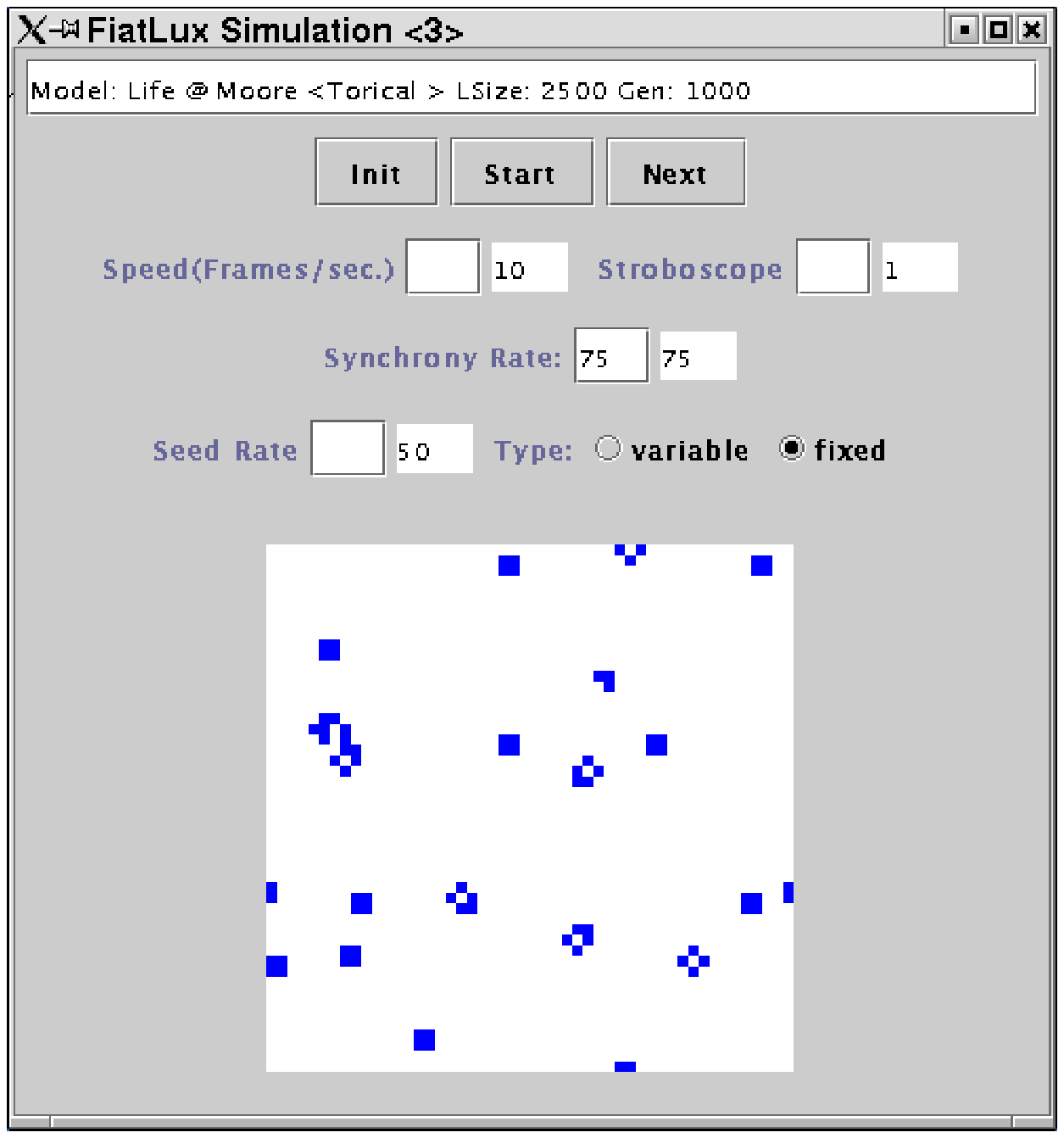}\\
 \Orbit{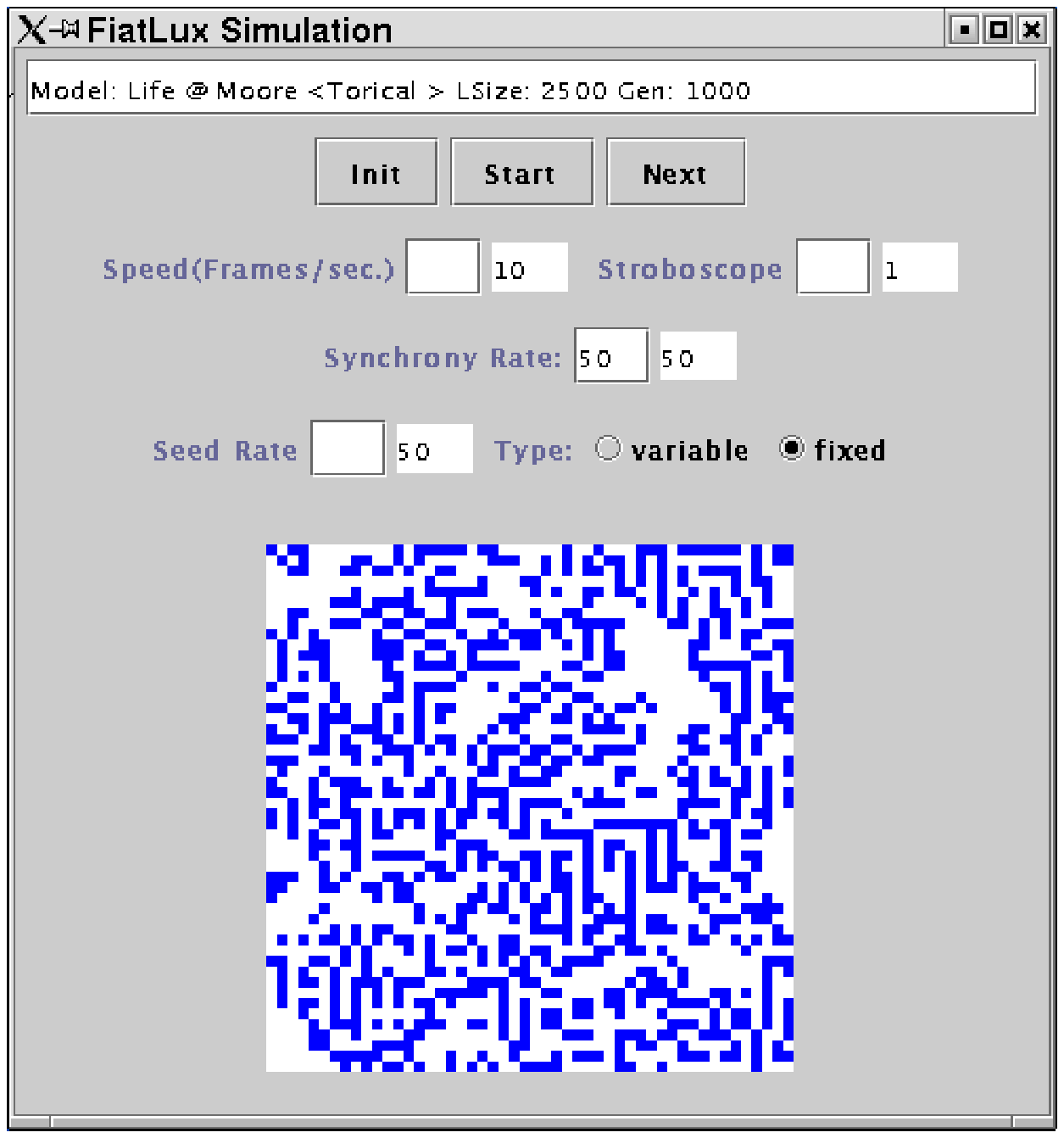} & \Orbit{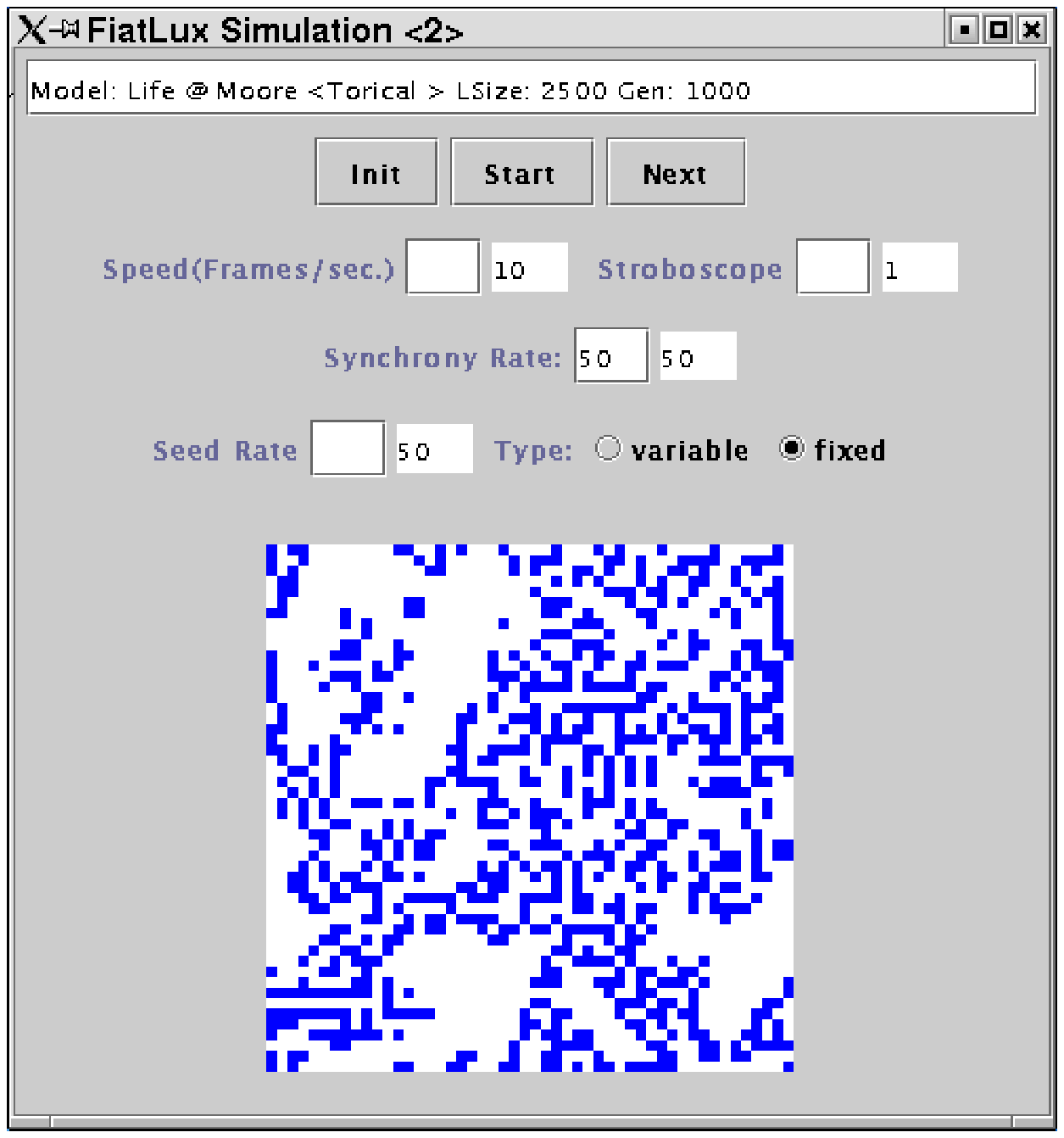} & \Orbit{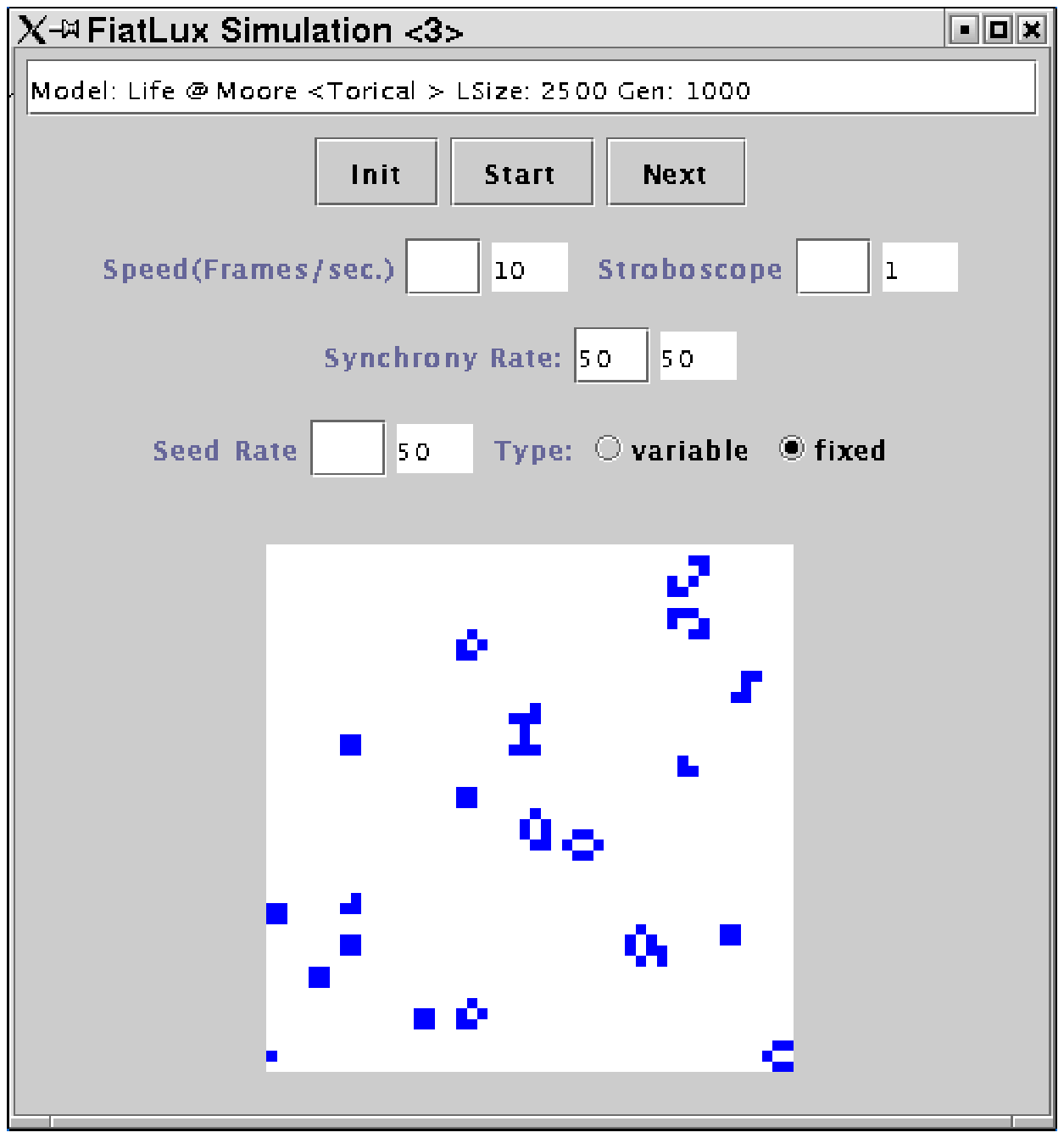}\\
    $ \er = 0 $      &  $ \er = 0.05 $     &   $ \er = 0.10 $    \\
\end{tabular}
\caption{\Life\ configurations for $ N=50 \times 50 $, after $ T=1000 $ time steps, starting from a random configuration of density $ \dini=0.5 $:
(up) $ \sr= 0.90 $  (middle) $ \sr= 0.75 $  (bottom) $ \sr= 0.50 $. The figure in the upper-left corner shows that the system is still in a transient mode.
}
\label{Fi:Life3}
\end{center}
\end{figure}

\subsection{Qualitative Observations} \label{Sec:Qualitative}

Figure \ref{Fi:Life1} shows that the behavior of \Life\ depends on the synchrony rate $ \sr $: a phase with labyrinthine shapes appears when $ \sr $ is lowered.
Bersini and Detours studied this phenomenon
and noticed that the asynchronous (sequential) updating of \Life\ was significantly different from the (classical) synchronous version
in that sense that a ``labyrinth phase'' (denoted by LP) appeared  (see \figref{Life1} below).
For small lattice dimensions, they observed the convergence of this phase to a fixed point and concluded 
that asynchrony had a stabilizing effect on \Life\ \cite{Ber94}.

The phase transition was then measured with precision by Blok and Bergersen, 
who used the final density (i.e, the fraction of 1's sites) as a means of quantifying the phase transition.
They measured the value $ \src $ for which the phase transition was to be observed and found $ \src = 0.91 $ \cite{Blo99}.
They showed that the type of phase transition is {\em continuous } (or a second-order transition): 
when $ \sr $ is decreased from $ \sr = 1.0 $ to $ \sr = \src $, 
no change is observed in terms of the values of the average density. 
When we have $ \sr < \src $ the ``labyrinth phase'' gradually appears and the average density starts increasing in a continuous way.
It is thus {\em the derivative} of the density that shows discontinuity rather than the function itself.

Figure \ref{Fi:ALife2} shows that the removal of links between cells does not qualitatively perturb 
the aspect of the final configurations attained.
So, according to the observation of steady states, synchronous \Life\ seems somehow robust to topology perturbations.
However, we also noticed that the transients are much shorter in presence of topology errors: 
for $ N= 50 \times 50 $, the order of magnitude of transients are $ T=1000 $ for $ \er =0 $ and $ T=100 $ for $ \er =0.1 $.

Figure \ref{Fi:Life3} shows what happens when both asynchronism and topology perturbations are added.
Rows of \figref{Life3} display the behavior with a fixed synchrony rate $ \sr $ and 
columns display the behavior with a fixed missing-link rate $ \er $.
We see that increasing topology errors from $ \er = 0 $ to $ \er = 0.05 $ makes the phase transition occur for a higher value of synchrony rate $ \src $.
With a further increase from $ \er = 0.05 $ to $ \er = 0.10 $, 
the phase transition cannot be observed any more, at least for the selected values of $ \sr$.

This demonstrates that both parameters $ \er $ and $ \sr $ control the phase transition 
between the ``inactive-sparse phase'' \cite{Hua03} and the ``labyrinth phase'' (LP).
The next section is devoted to quantitatively measure the interplay of these two control parameters.

\subsection{Quantitative Approach} \label{Sec:ExperimentalData}

\BF 

\InsertBiFig{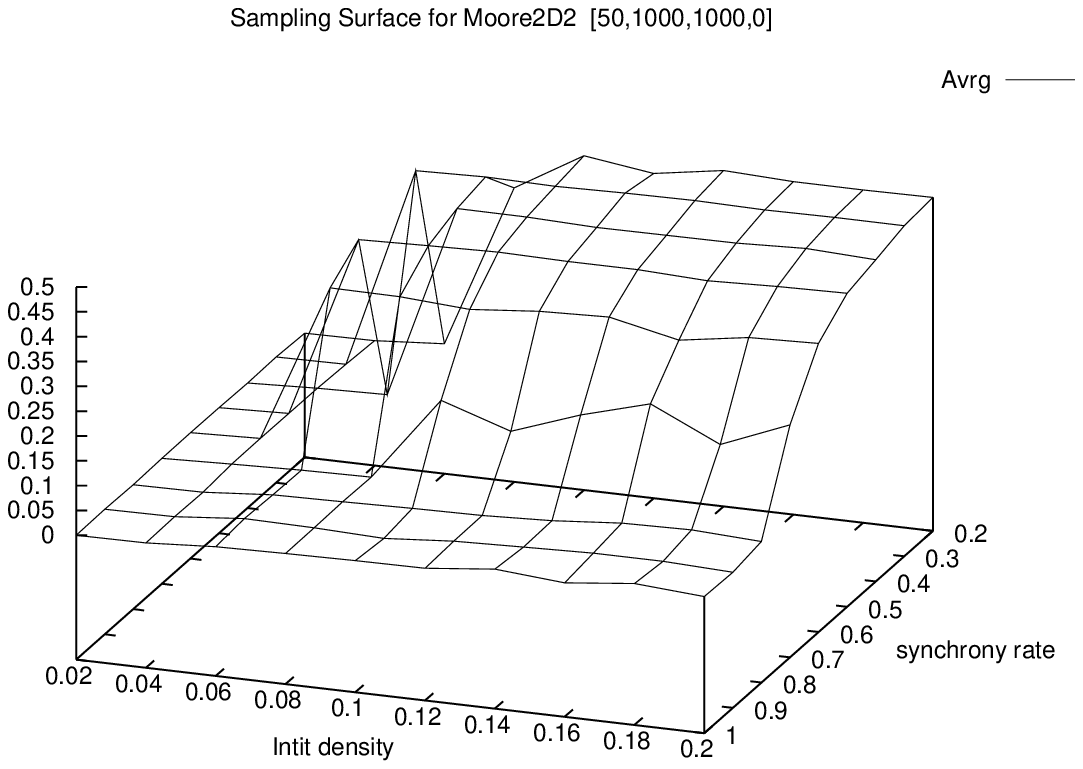}{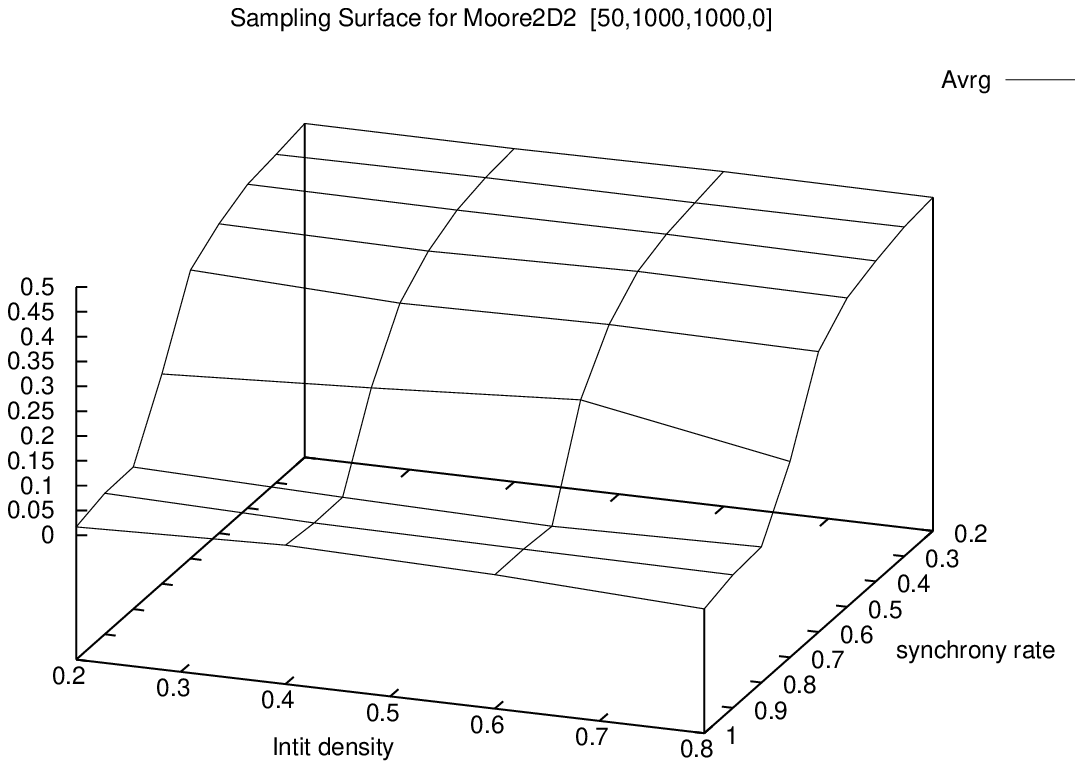}{\SzA}
\InsertBiFig{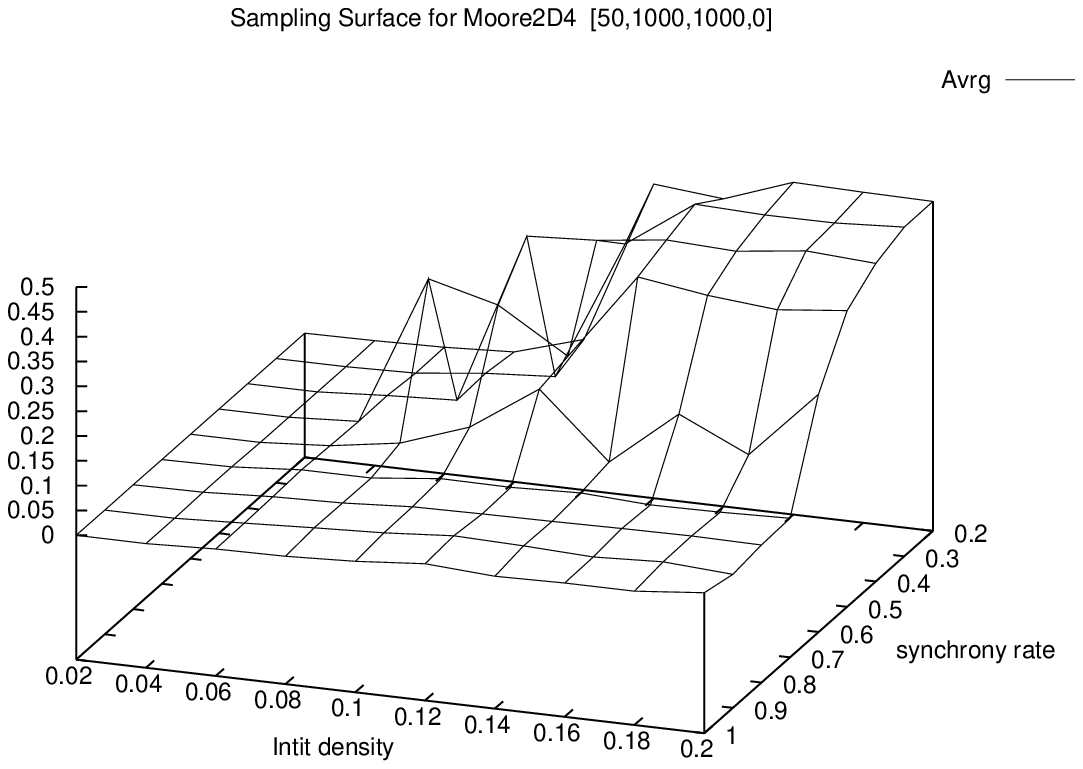}{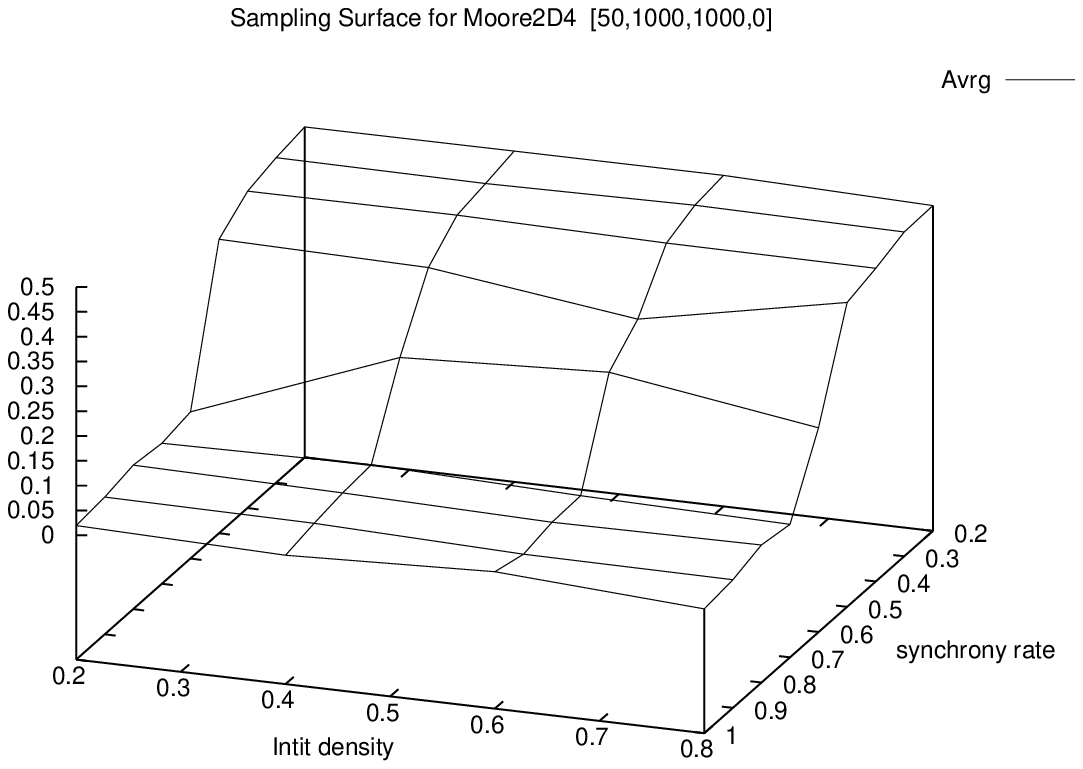}{\SzA}
\InsertBiFig{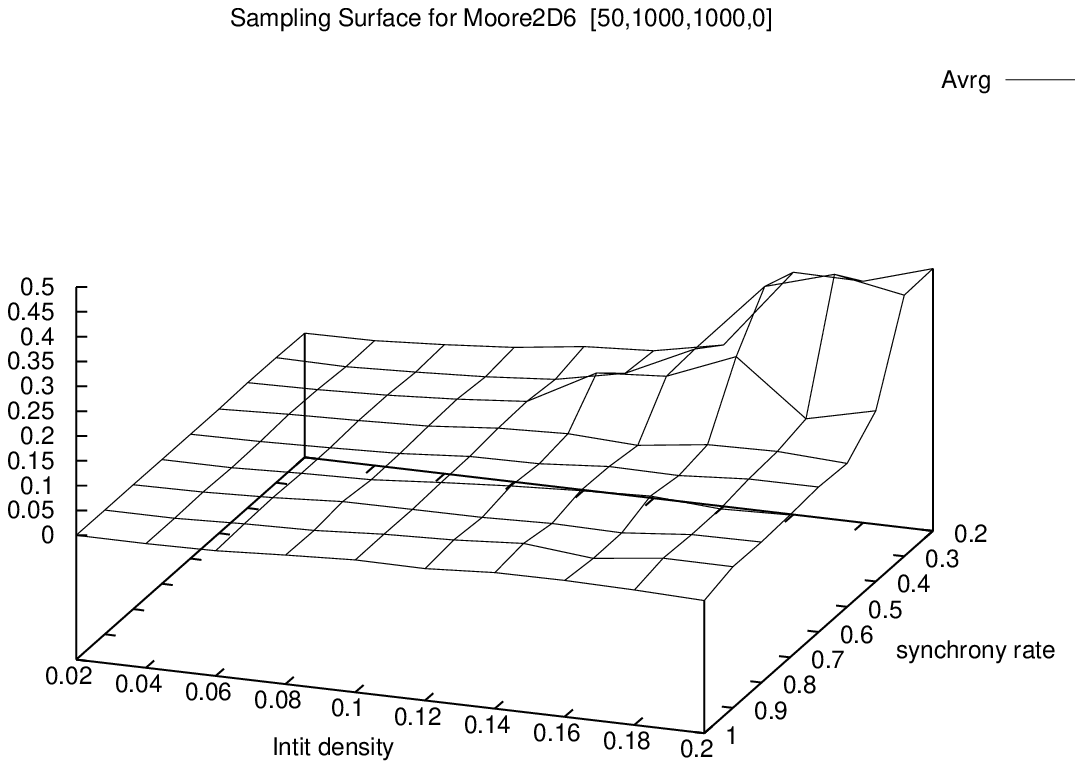}{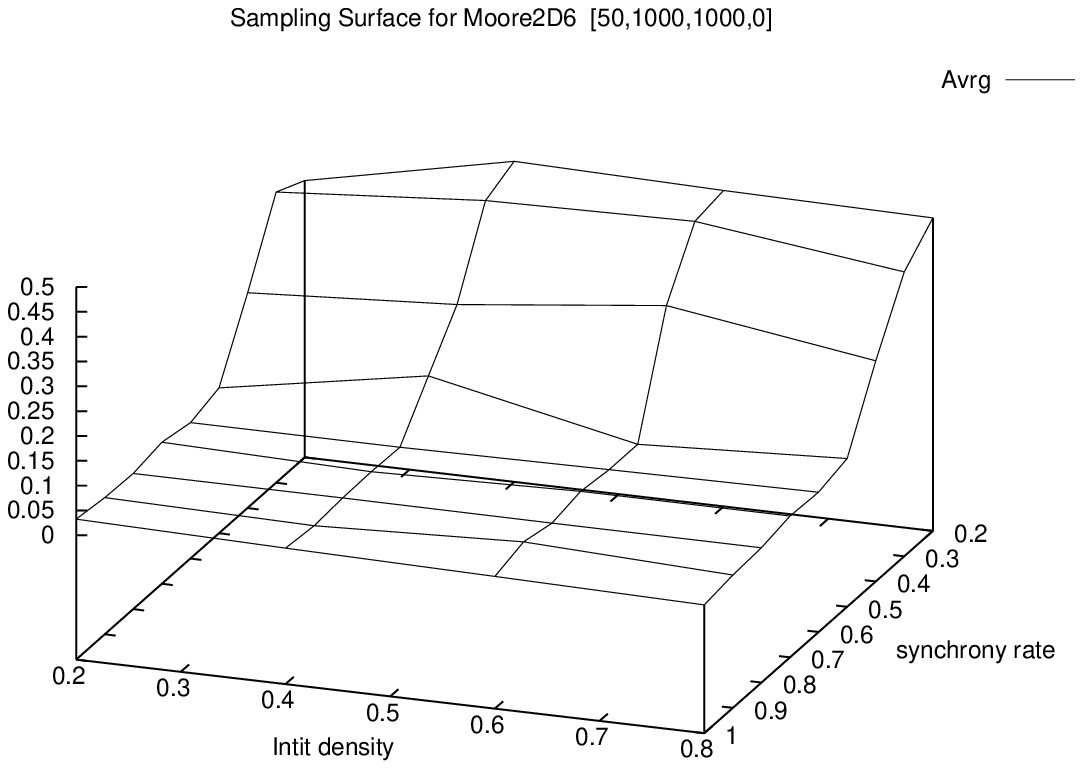}{\SzA}
\InsertBiFig{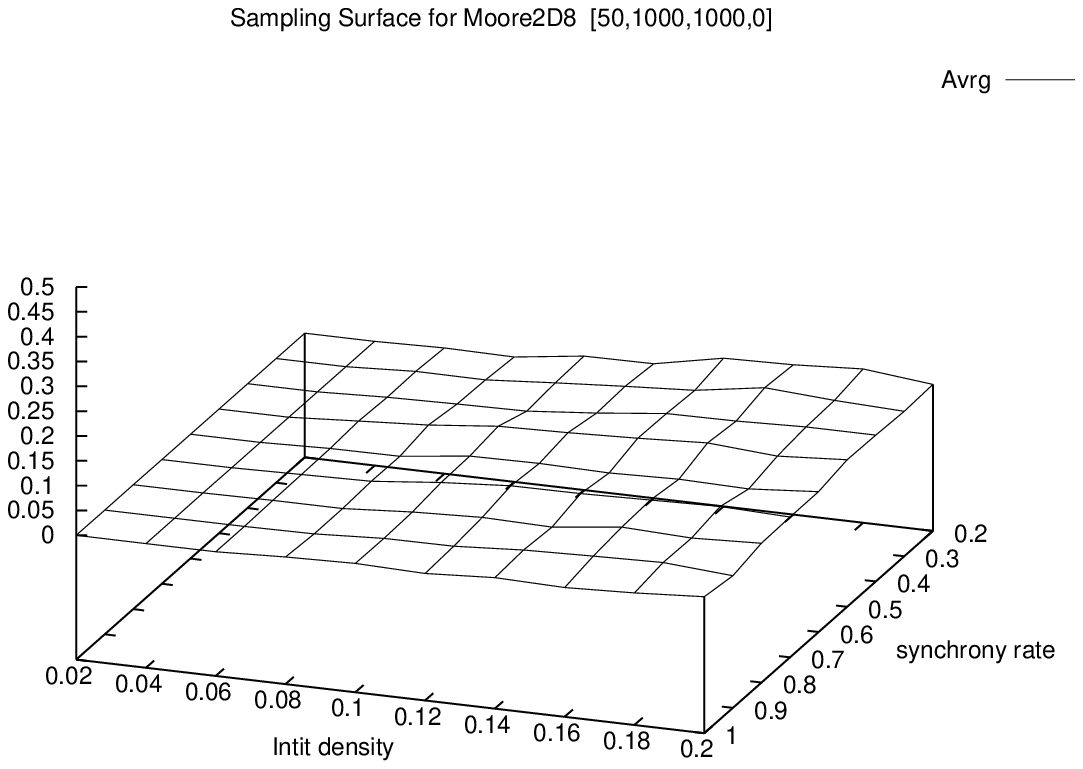}{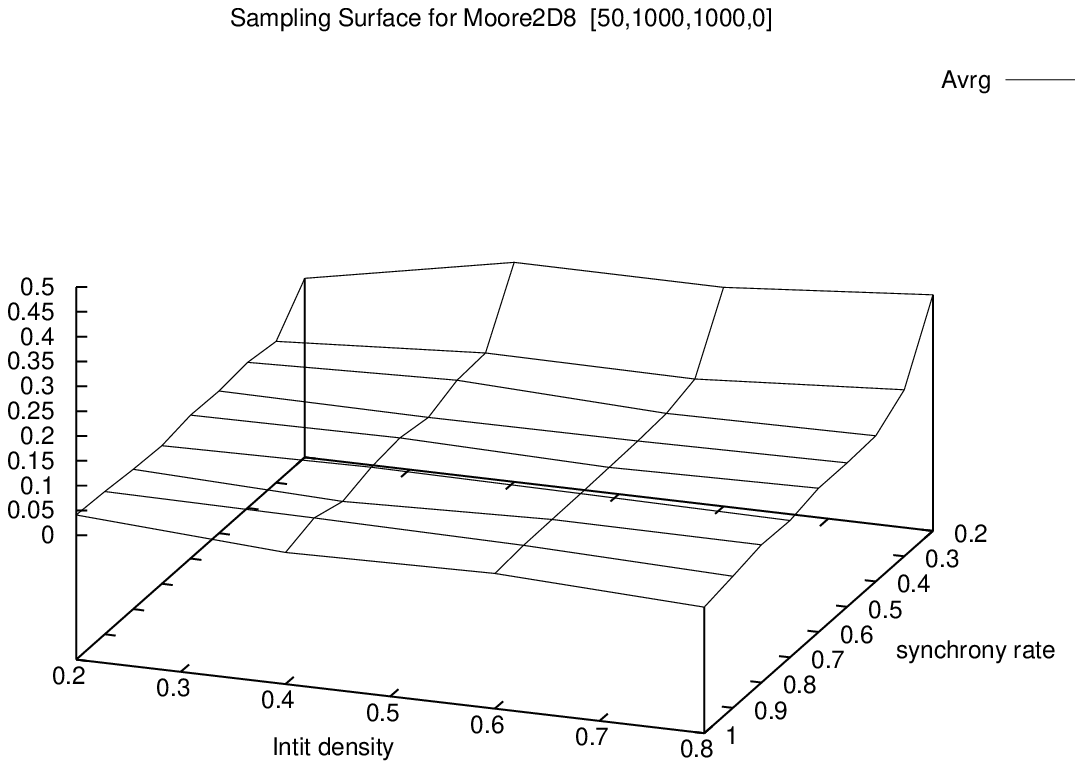}{\SzA}
\InsertBiFig{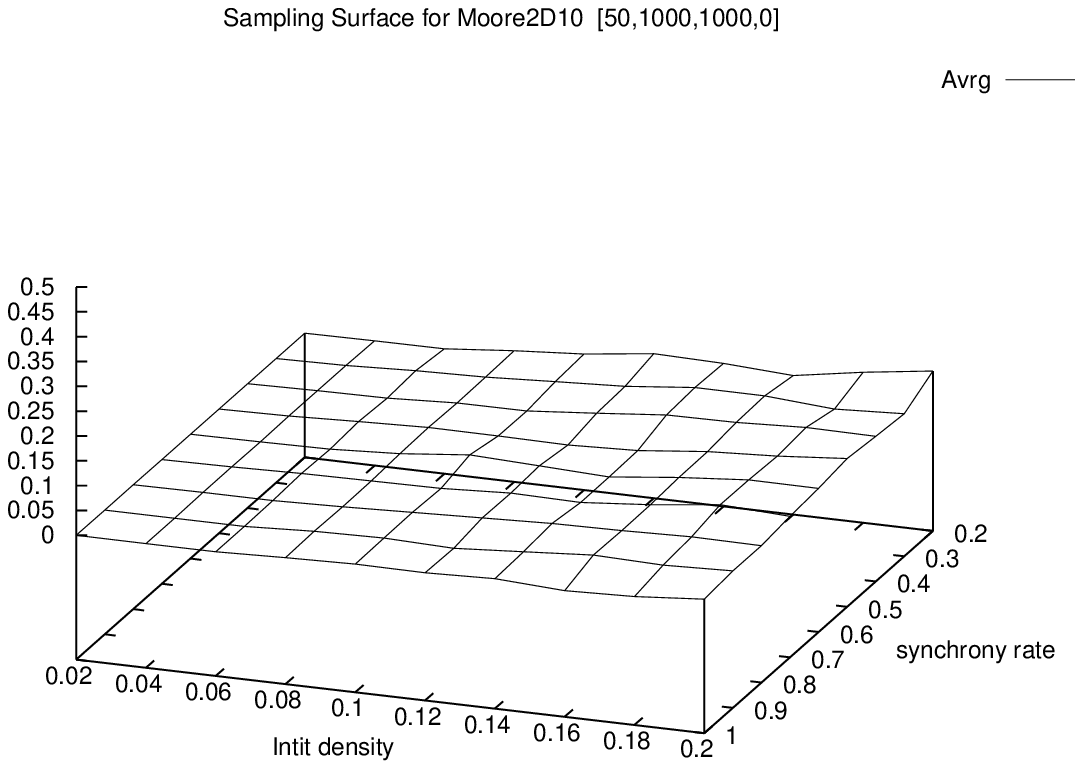}{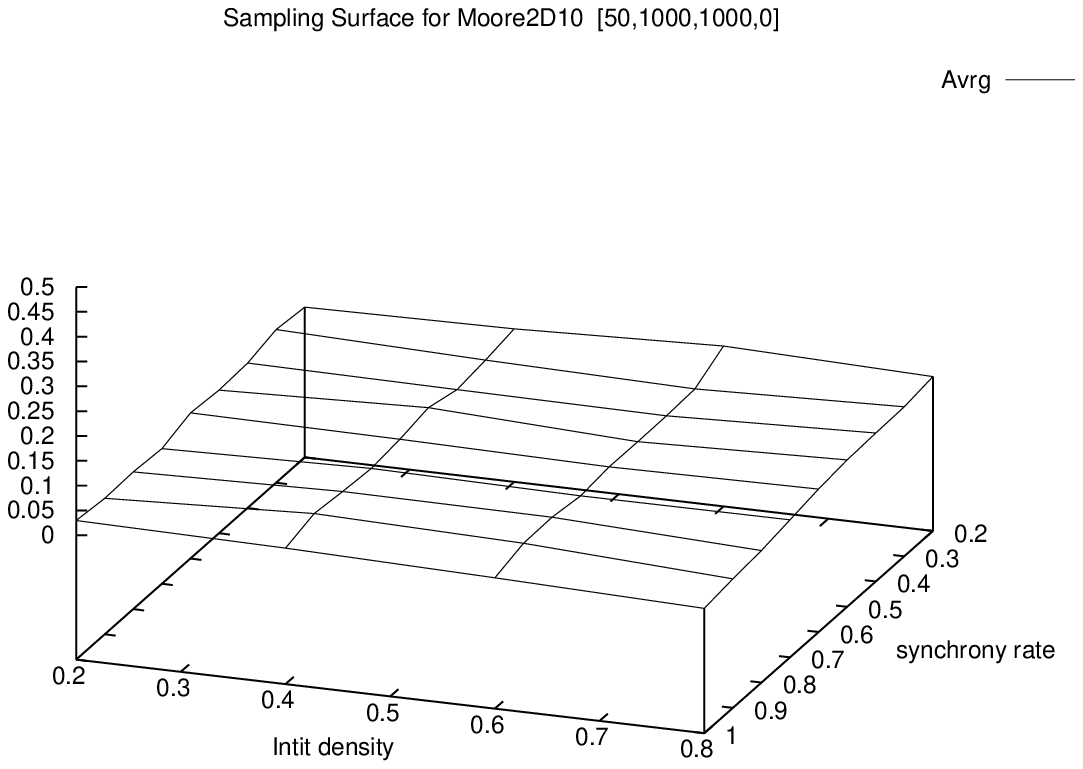}{\SzA}

\caption{Sampling surfaces for $ \er= [0 \cdots 0.10] $, $ N= 50  \times  50 $, $ \Tt = 1000 $ , $ \Ts = 1000 $. 
The left column has a different range for $ \dini$ to focus on the behavior for small initial densities.}
\label{Fi:SamplingSurfaces}
\EF

To detect the apparition of the labyrinth phase (LP), we need to look at the configurations by eye or to choose an appropriate macroscopic measure.
Clearly, a configuration in LP contains much more $\stone $'s than a configuration in the ``inactive-sparse phase'' (\cite{Hua03}).
This leads us to quantify the change of behaviour using the measurement of the ``steady-state density'' 
(i.e. the average density after a transient time has elapsed).
This method has been chosen by various authors (e.g. Blok and Bergersen \cite{Blo99}) and
it has been applied to exhaustively study both the dynamics \cite{Fates03} and the robustness to asynchronism \cite{Fat04} of 
one dimensional elementary cellular automata.

We define the steady-state density $ \rho(\dini,\sr) $ using the sampling algorithm defined in \cite{Fat04}:
Starting from a random initial configuration constructed with a Bernoulli process of parameter $ \dini $, 
we let the ACA evolve with a synchrony rate $ \sr $ during a transient time $ \Tt $
then we measure the value of the density during a sampling time $ \Ts $.
The value of $ \rho $ is the average of the sampled densities.

The sampling operation results in the definition of a function $ \rho ( \dini, \sr ) $ 
that can be represented in the form of a ``sampling surface''. 
This surface contains part of the information on how the behaviour of a CA is affected by asynchronism.
Figure \ref{Fi:SamplingSurfaces} shows the experimental results obtained for $ N= 50  \times  50 $, $ \Tt = 1000 $ , $ \Ts = 1000 $.
The transient time $ \Tt $ is chosen according to the observations made in \cite{Bag91} where equivalent transient times were found 
for greater lattice sizes.

Let us first look at what happens for $ \dini \in [0.2,0.8] $ (right column of \figref{SamplingSurfaces}).
The invariance of the surface relatively to the $ \dini $-axis 
shows that the macroscopic behavior of \Life\ does not depend on the value of this parameter within this range.
The upper right corner of \figref{SamplingSurfaces} shows that for $ \er = 0 $ (regular topology), 
the phase transition occurs for $ \src \eq 0.90 $ as expected \cite{Blo99}.
However, when $ \er $ increases, experiments show that $ \src $ also decreases. 
This means that the settlement of LP becomes more difficult as links are removed; 
this can be interpreted as an increase of the robustness to asynchrony. 
We can observe that for $ \er = 0.10 $ , the surface is flat and horizontal, 
which means that the behavior is not anymore perturbed by asynchronism (at least if we consider our observation function).

The left column of \figref{SamplingSurfaces} shows the behavior of \Life\  for $ \dini \in [0,0.2] $.
We observe two different abrupt change of behaviors. 
On the one hand, there is a value of $ \dinic $ which separates the ``inactive-sparse phase'' and LP. 
On the other hand, the value of $ \dinic $ increases as $ \er $ increases. 
This means that LP becomes more difficult to reach when links are removed; which again can be interpreted as a gain of robustness.

Experiments were held for various lattice sizes and allowed to control that the sampling surface aspect was stable with $ N $; 
we however observed that when $ \er $ and $ \sr $ are fixed, the value of $ \dinic $ is a decreasing function of $ N $.

All the previous phenomena may be the consequence of multiple intricate factors. In the next section, we study the evolution
of so called ``micro-configurations'' put in an empty array and propose a first hypothesis in the direction of understanding these
behaviors.

\section{Micro-configurations Analysis} \label{Sec:Analysis}

\subsection{Experiments}

The observation of the settlement of LP shows that it can develop from very localized parts of the lattice 
and then spreads across the lattice until it fills it totally. By analogy with a crystal formation, 
we call ``germs'' these particular configurations that have the possibility to give birth to an invasive LP.
We investigate the existence of germs by performing an exhaustive study of the potential evolution of 
{\em micro-configurations}, i.e. $ 3 \times 3 $ configurations that are placed
in an empty array. There are 512 such configurations and we experimentally quantify, for each one, 
the probability that a it becomes a germ. 
Our goal is to infer the behavior of the whole structure from the evolution of these micro-configurations.

Setting the synchrony rate to $ \sr = 0.5 $, we used the following algorithm:\\
For every micro-configuration $ i \in I $, (a) we initialize the lattice $ N  \times  N$ with $ i$,
and (b) we let the CA evolve until it reaches a fixed point or until it reaches LP.
We repeat $ S = 1000 $ times operations (a) and (b) for the same initial micro-configuration but for a different update histories.
We consider that the CA has reached LP if the density is greater of equal than a limit density $ \dLP = 0.1 $.
Indeed, we observed that if the CA was able to multiply the number of $\stone$'s 
from the micro-configuration to a constant ratio of the lattice, 
then it will almost surely continue to invade the whole lattice 
and, asymptotically, reach LP. 
We experimentally obtain the probability $ \Pgerm[i] $ that a configuration $ i $ is a germ.
Grouping micro-configurations by the number $ k$ of $\stone$'s they contain, 
we obtain an array with 9 entries $ \Pgerm(k), k \in [0,9] $, displayed in Table \ref{Tab:probaTopo}.

Results show that for $ k < 3, \Pgerm(k) = 0 $, 
which means that all such micro-configurations always tend to extinction. 
For $ n \geq 3$, the probability to reach LP increases as $ n $ increases.

\begin{table}[tbp]
	\caption{Probability $ \Pgerm(k) $ for a micro-configuration to be a germ (in \%) as a function of the number $ k $ of $\stone$'s for different missing-link rates $\er $ (in \%).}	
	\begin{center}
	\include{Histogram}
	\end{center}
	\label{Tab:probaTopo}
\end{table}

\begin{figure}[htbp]
	\InsertFig{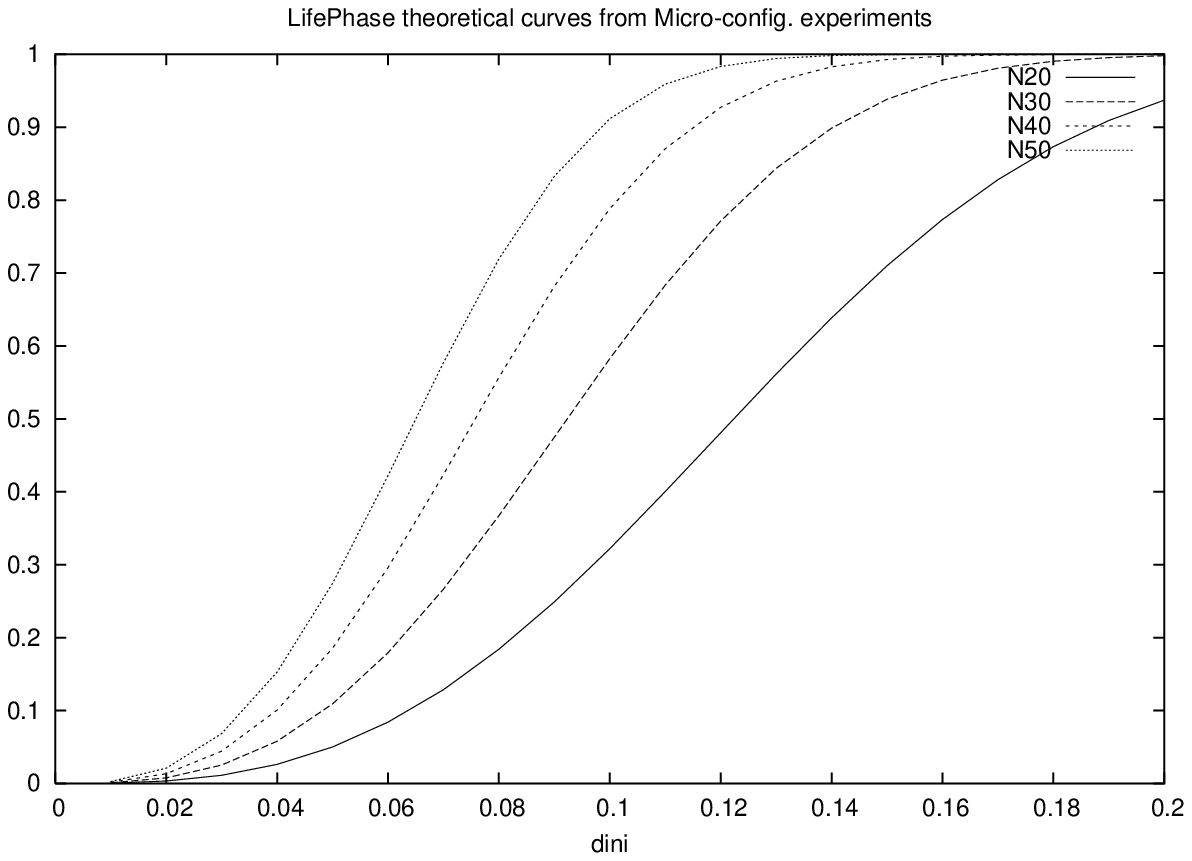}{\szEndFig}		\InsertFig{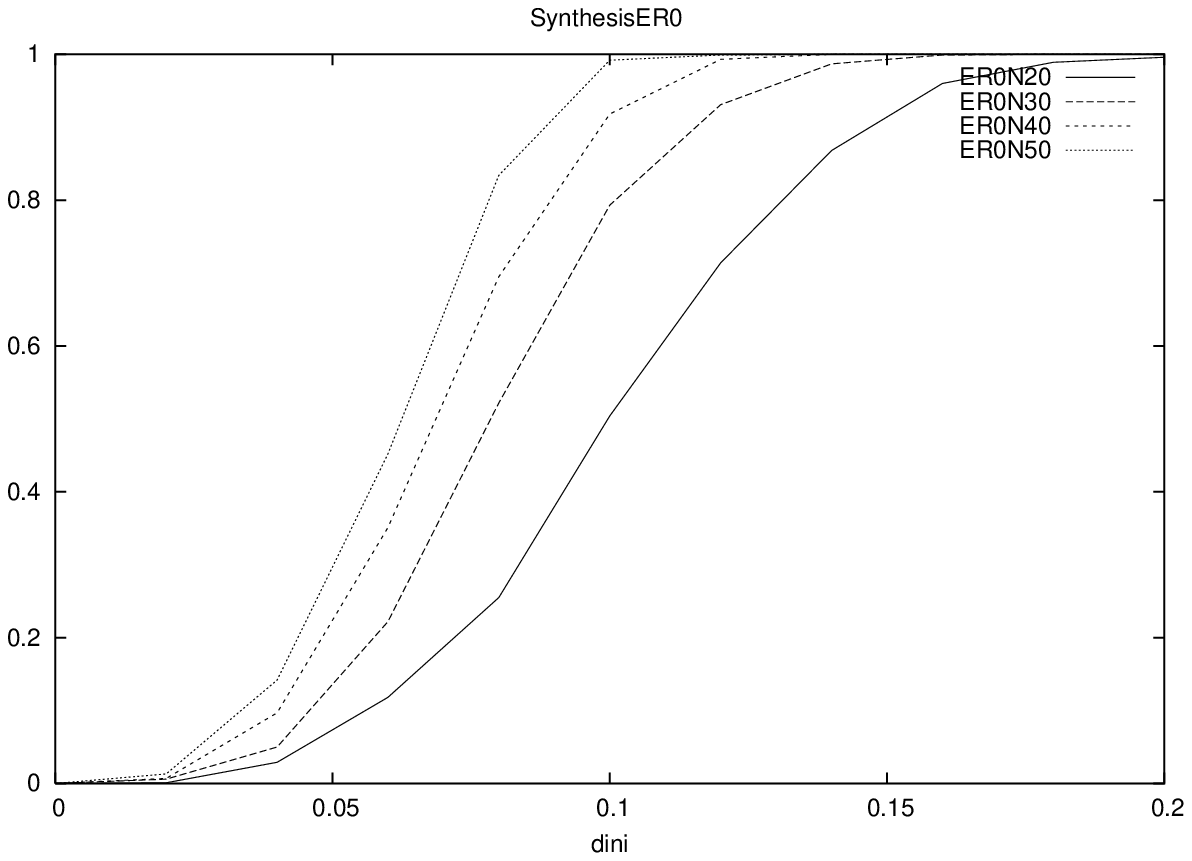}{\szEndFig} \\		
	\InsertFig{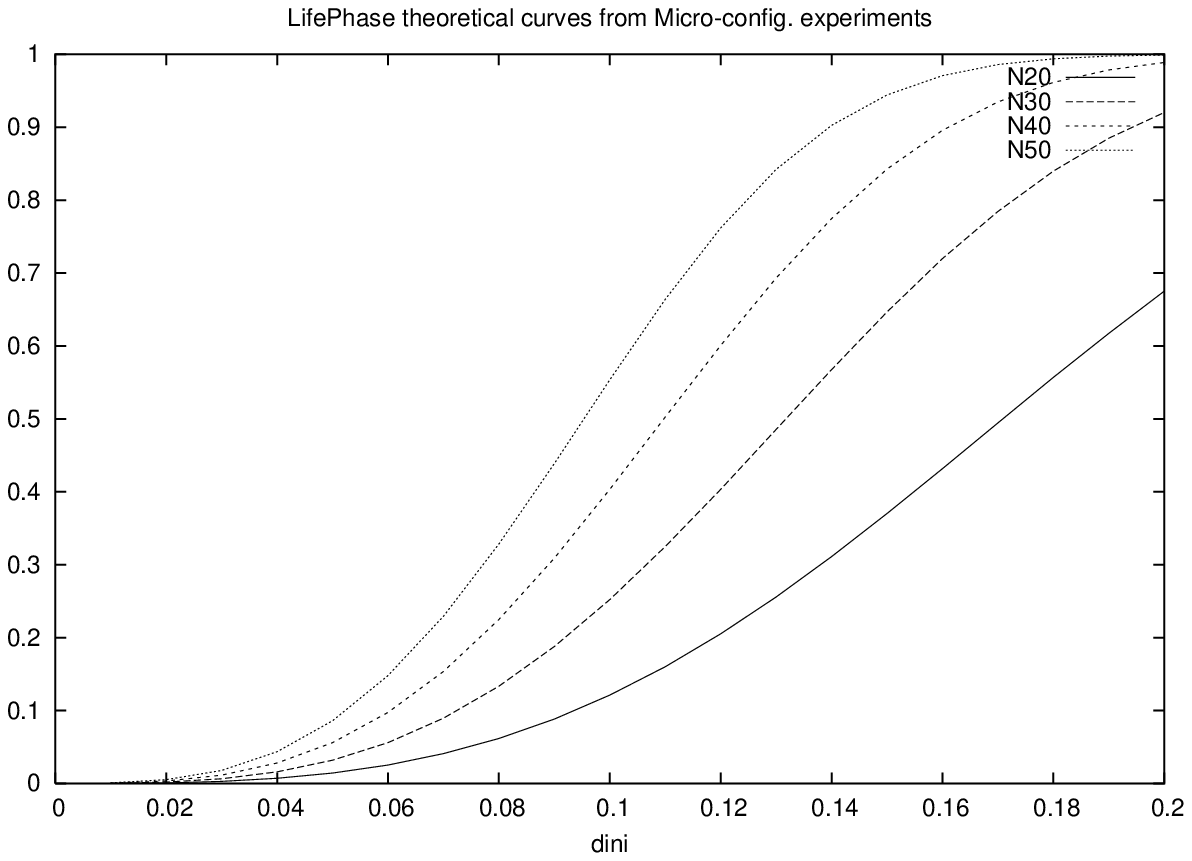}{\szEndFig}		\InsertFig{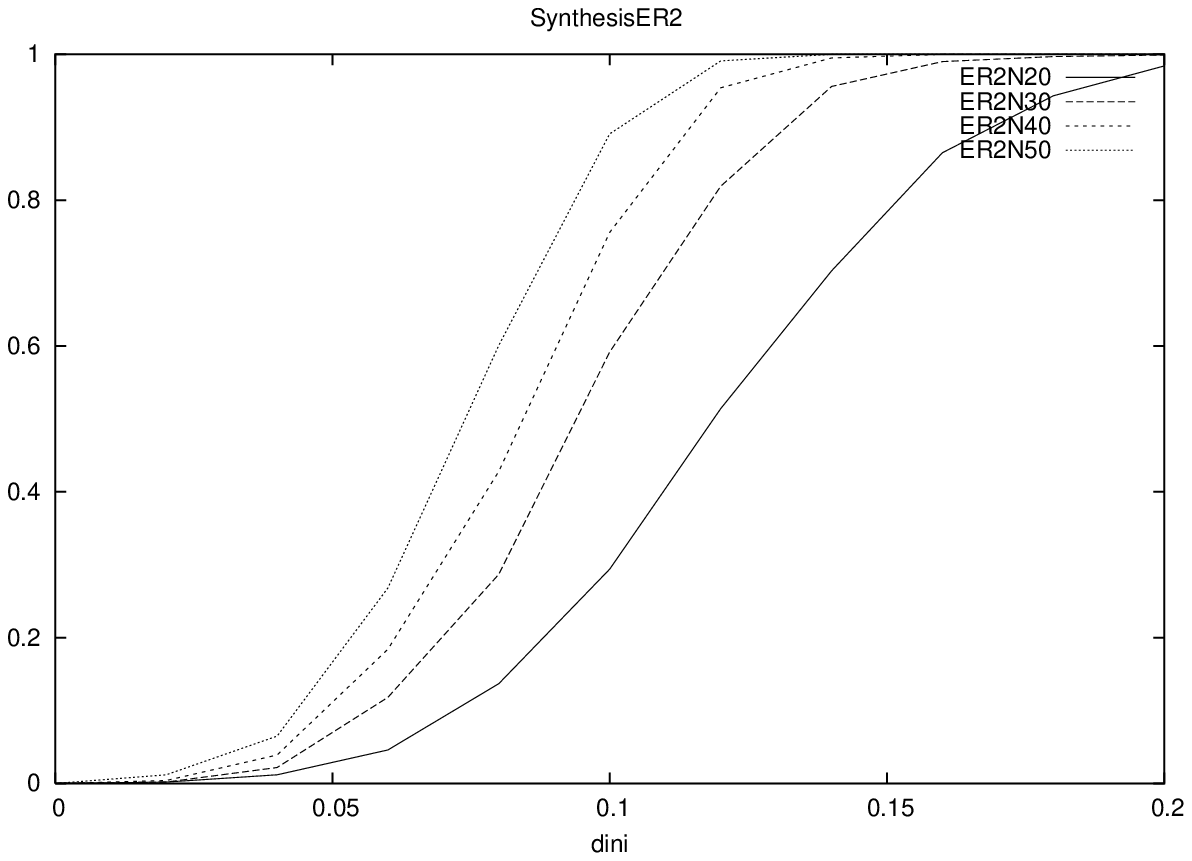}{\szEndFig} \\			
	\InsertFig{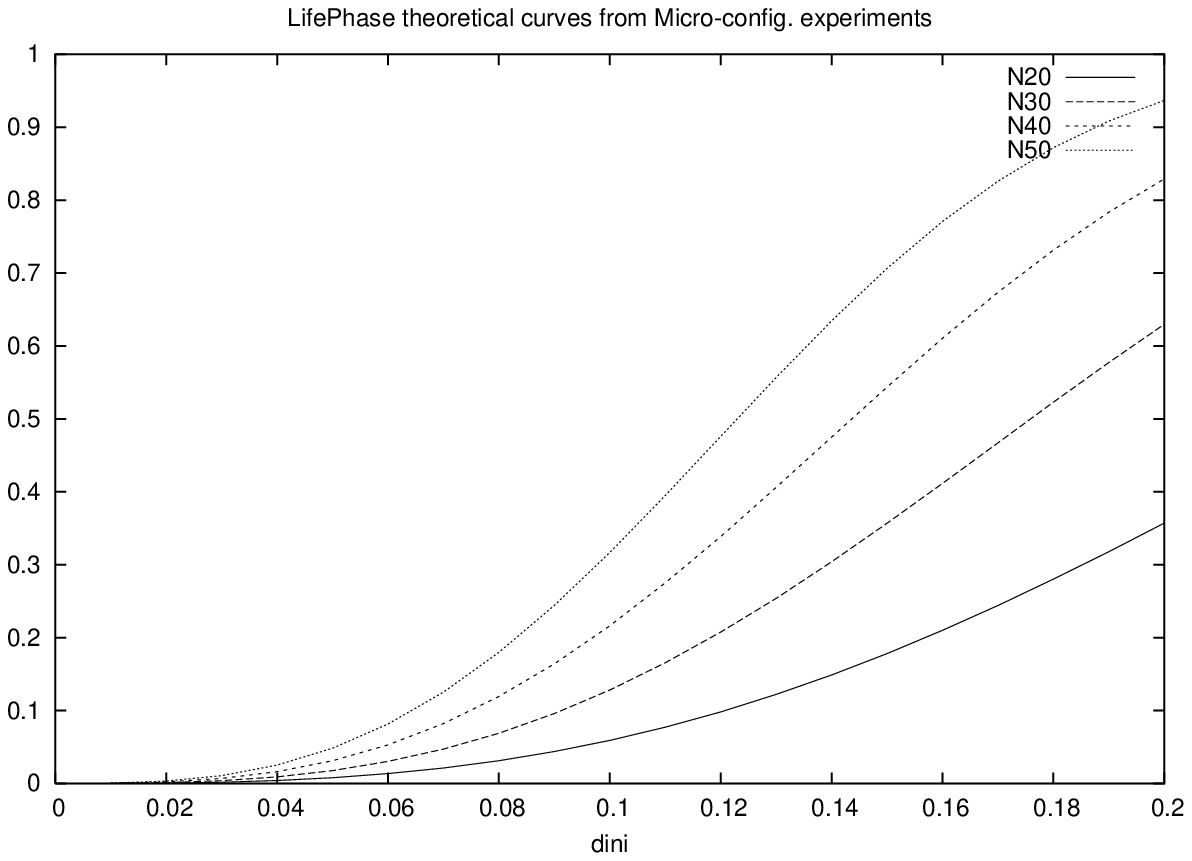}{\szEndFig}		\InsertFig{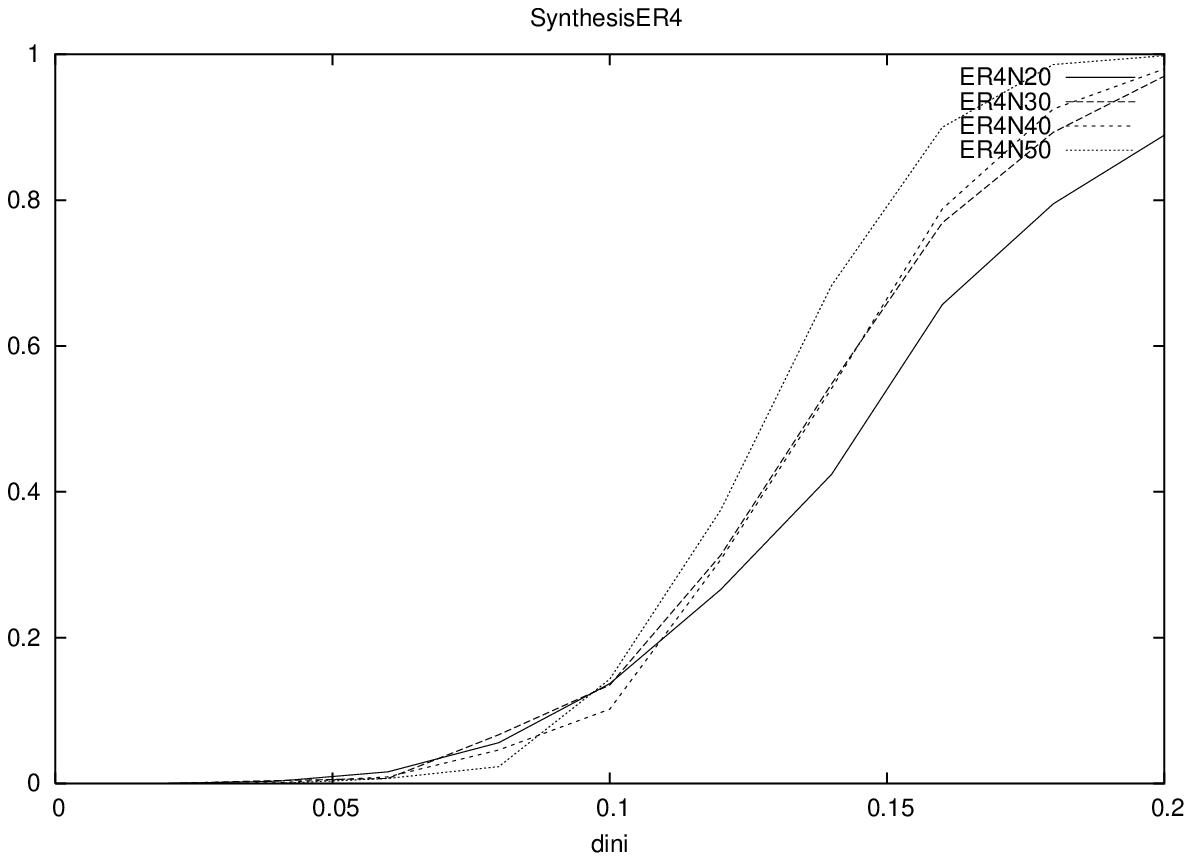}{\szEndFig} \\			

\caption{Probability to reach LP as a function of $ \dini $ for different lattice sizes $ N \in \set{20, 30, 40, 50} $ and different missing-link rates: 
$ \er = 0 $, $ \er= 0.02 $, $ \er= 0.04 $ from top to bottom. Left column shows theoretical curves calculated with the independent-germ hypothesis; the right column shows the actual measurements.}
	\label{Fi:ProbaLP}
\end{figure}

\subsection{Inferring Some Aspects of the Global Behavior}

Our idea is to use the previous results about germs to extend them to a description of the global behavior of \Life.
Unfortunately, we are not able to do that in the exact way. 
In a first approximation, let us assume that we can infer this global behavior 
by approximating the probability $ \PLP $ for a uniformly initialized system to reach LP using an ``independent-germ hypothesis'': 
interactions between potential germs are neglected and we assume that 
LP is reached if and only if there is at least one cell that gives birth to LP.
This results in the application of formula :
$ \PLP = 1 - (1 - \PLPone )^{N*N} $
where $ \PLPone $ is the probability that one cell gives birth to LP.
We have $ \PLPone = \sum_{k=0}^9 \Pgerm(k) \cdot P_{\mathrm{app}}(k,\dini) $,
where $ P_{\mathrm{app}} (k,\dini) $ is the probability that a micro-configuration 
initialized randomly with $ \dini $ contains  $ k $ $\stone$'s. 
It is simply obtained by applying the binomial formula:
$ P_{\mathrm{app}}(k,\dini) = \binom{9}{k} {\dini}^k (1-\dini)^{9-k} $.

Calculated and experimental values of $ \PLP(\dini,\er) $ are given in \figref{ProbaLP} 
and show that this assumption is justified in a first approximation
even if the predictions seem more accurate for small values of $ \er $.

The germ hypothesis allow us to understand better some of the observed behavior.
Let us first consider the abrupt change of behavior observed for $\dini \in [0,0.20] $: 
this can come from the fact that
as $ \dini $ increases the probability to observe a micro-configuration that contains more $\stone$'s increases thus
increasing the probability to find a germ in the initial configuration.
We can also understand with this point of view the invariance of the sampling surfaces in the $ \dini$-axis with $ \dini > 0.20 $ 
by the fact, observed in \figref{ProbaLP} that in this case $ \PLP \eq 1 $, 
that is the ``labyrinth phase'' always appears.
In the same way, the shift of $ \dinic $ observed in \figref{SamplingSurfaces} when varying $\er$ can be explained
by the looking at the variations of $ \Pgerm(k) $ with $ \er $ : 
we see that all probabilities to reach LP decrease when $ \er $ increases.
Finally, we are able to qualitatively predict the scaling of $ \dinic $ with the lattice size $ N $ 
from the plots the function $ \PLP(\dini, \er) $ :
when $ \dini $ and $ \PLP $ are relatively small (i.e., $ \PLP < 0.1 $), we have a linear scaling of $ \PLP  $ with $ N² $;
whereas as $ \PLP $ is close to saturation, there tends to be no variations with $ N $.
\section{Conclusion}

Experiments have shown that \Life's transition from an ``inactive-sparse phase'' to a ``labyrinth phase'' (LP) 
is a continuous phase transition dependent on the synchrony rate $ \sr $ and 
whose critical value $ \src $ is controlled by the missing-link rate $ \er $.
As the topology was perturbed (i.e., when $ \er $ increased), the inactive-sparse phase domain extends
while the LP domain shrinks.
The abrupt change in behavior according to the values of initial density $ \dini $ 
was interpreted with the hypothesis that the settlement of LP results from the development of ``germs'', 
i.e. small configurations that are able to ``colonize'' the whole lattice.
The study of the evolution of potential germs from micro-configurations allowed us to start to understand the observations 
and to give some predictions on the probability to reach LP starting from a random configuration.
One interesting question is to examine whether these observations hold for a large class of CA or 
if they are somehow related to the computational universality of \Life.

{
\bibliographystyle{splncs}
\bibliography{CAbiblio.bib}
}

\end{document}

%% file: commands.tex
\usepackage{amsmath}
\usepackage{amsfonts}
\usepackage{amssymb}

\newcommand{\Z}{\mathbb{Z}} 			

\newcommand{\U}{{\cal L}} 			

\newcommand{\eq}{\thicksim}			
\newcommand{\set}[1]{\{#1\}}

\newcommand{\dini}{d_{\mathrm{ini}}}
\newcommand{\dinic}{\dini^\mathrm{c}}	
\newcommand{\sr}{{\alpha}}
\newcommand{\src}{{\alpha}_{\mathrm{c}}}
\newcommand{\er}{{\epsilon^-}}
\newcommand{\Pgerm}{{P}_{\mathrm{germ}}}
\newcommand{\PLP}{{P}_{\mathrm{LP}}}
\newcommand{\PLPone}{{P}_{\mathrm{LP1}}}
\newcommand{\dLP}{{d}_{\infty}}

\newcommand{\Tt}{ T_{\mathrm{transient}} }
\newcommand{\Ts}{ T_{\mathrm{sampling}} }

\newcommand{\Word}[1]{\texttt{#1}}
\newcommand{\stzero}{\Word{0}}
\newcommand{\stone}{\Word{1}}

\newcommand{\DebutTableau}[1]{
\begin{center}
\begin{tabular}{#1}
\hline
}

\newcommand{\FinTableau}{
\hline
\end{tabular}
\end{center}
}

\usepackage{epsfig}

\usepackage{subfigure}
\newcommand{\BF}{\begin{figure}[htbp] \begin{center} }
\newcommand{\EF}{\end{center} \end{figure} }

\newcommand{\InsertFig}[2]{
	\centering	\epsfig{file=#1.eps, width=#2 cm}
}

\newcommand{\InsertBiFig}[3]{	
	 	\InsertFig{#1}{#3}
 		\InsertFig{#2}{#3}	
}

\newcommand{\Orbit}[1]{\epsfig{file=#1.eps, width=3 cm, clip=, bbllx=3.3cm, bblly=1.1cm, bburx=10.2 cm, bbury= 7.7cm}}

\newcommand{\figref}[1]{Fig. \ref{Fi:#1}}

\newcommand{\BeginList}{\begin{itemize}}
\newcommand{\EndList}{\end{itemize}}

\newcommand{\BQ}{\begin{quote}}
\newcommand{\EQ}{\end{quote}}

%% file: Histogram.tex
\newcolumntype{Y}{>{\centering\arraybackslash}X}
\begin{tabularx}{9cm}{|c|Y|Y|Y|Y|Y|Y|Y|Y|Y|Y|}  
 \hline 
k & 0 & 1 & 2 & 3 & 4 & 5 & 6 & 7 & 8 & 9 \\
\hline
$ \er =0 $ &  0 & 0 & 0 & 1.28 & 4.34 & 7.88 & 14.52 & 21.76 & 29.92 & 40.90 \\
$ \er =2  $ & 0 & 0 & 0 & 0.28 & 2.17 & 3.93 & 7.22 & 11.18 & 15.34 & 21.70 \\
$ \er =4  $ & 0 & 0 & 0 & 0.19 & 0.72 & 1.24 & 2.40 & 3.96 & 5.52 & 8.00 \\
$ \er =6  $ & 0 & 0 & 0 & 0.02 & 0.06 & 0.09 & 0.24 & 0.34 & 0.63 & 1.30 \\
 \hline  
\end{tabularx}